\documentclass[twocolumn]{IEEEtran}
\usepackage{amsmath, amsfonts, amssymb}
\usepackage{graphicx,epsfig}
\usepackage{color}
\numberwithin{equation}{section}

\usepackage{dsfont}
\usepackage{hyperref}
\usepackage{array}
\usepackage{booktabs}
\graphicspath{{images/}}

\usepackage{amsthm}
\theoremstyle{plain}\newtheorem{remark}{Remark}[section]
\theoremstyle{definition}

\begin{document}

\title{On the extraction of instantaneous frequencies from ridges in time-frequency representations of signals}

\author{D. Iatsenko, P. V. E. McClintock, A. Stefanovska}

\maketitle

\begin{abstract}
The extraction of oscillatory components and their properties from different time-frequency representations, such as windowed Fourier transform and wavelet transform, is an important topic in signal processing. The first step in this procedure is to find an appropriate ridge curve: a sequence of amplitude peak positions (ridge points), corresponding to the component of interest. This is not a trivial issue, and the optimal method for extraction is still not settled or agreed. We discuss and develop procedures that can be used for this task and compare their performance on both simulated and real data. In particular, we propose a method which, in contrast to many other approaches, is highly adaptive so that it does not need any parameter adjustment for the signal to be analysed. Being based on dynamic path optimization and fixed point iteration, the method is very fast, and its superior accuracy is also demonstrated. In addition, we investigate the advantages and drawbacks that synchrosqueezing offers in relation to curve extraction. The codes used in this work are freely available for download.
\end{abstract}


\section{Introduction}

Separation of amplitude and frequency-modulated components (AM/FM components) in a given signal, and estimation of their instantaneous characteristics, is a classical problem of signal analysis. It can be approached by projecting the signal onto the time-frequency plane, on which the changes of its spectral content can be followed in time. Such projections are called time-frequency representations (TFRs), with their typical examples being the windowed Fourier transform (WFT) and the wavelet transform (WT). If the construction of the TFR is well-matched to the signal's structure, then each AM/FM component will appear as a ``curve'' in the time-frequency plane, formed by a unique sequence of TFR amplitude peaks -- ridge points. Based on properties of these curves, one can estimate the time-varying characteristics of the corresponding components (such as amplitude, phase and instantaneous frequency), an idea that was first expressed in \cite{Delprat:92} (for a discussion of different reconstruction methods and their performance, see \cite{Iatsenko:tfr}).

However, to estimate the parameters of the component in this way, one first needs to extract its associated ridge curve, i.e.\ find the corresponding peak sequence. This is not a trivial issue, since in real cases there are often many peaks in the TFR amplitude at each time, and their number often varies. In such circumstances it can be unclear which peak corresponds to which component, and which are just noise-induced artifacts.

In the present paper, we concentrate solely on the problem of the ridge curve identification, which is of great importance in time-frequency signal processing. Thus, ridge analysis is widely used for e.g.\ machine fault diagnosis \cite{Zhang:03}, fringe pattern analysis \cite{Zhong:05}, studies of cardiovascular dynamics \cite{Iatsenko:cardio} and system classification \cite{Suprunenko:13,Staszewski:98}. Although curve extraction has been addressed explicitly in the past \cite{Carmona:97,Carmona:99,Iatsenko:cardio,Barros:01,Thakur:13}, there seems to be no agreement as to the optimal procedure to be used for this task. Here we discuss and generalize some existing algorithms, present new ones, and compare their performance. We end up with a method that is accurate and of universal applicability, so that it works well for a large class of signals and, in most cases, does not require adjustment by the user; this is the main contribution of the work. The effects of synchrosqueezing \cite{Daubechies:11,Thakur:11,Thakur:13,Auger:13} on curve extraction are also studied.

The plan of the work is as follows. After reviewing the background and notation in Sec.\ \ref{sec:background}, we discuss different schemes for curve extraction in Sec.\ \ref{sec:schemes}. In Sec.\ \ref{sec:comparison} we compare the performance of these schemes, while the advantages and drawbacks of synchrosqueezing in relation to curve extraction are studied in Sec.\ \ref{sec:ss}, and the limitations of the proposed methods are discussed in Sec.\ \ref{sec:limitations}. We draw conclusions and summarize the work in Sec.\ \ref{sec:conclusions}. A dynamic programming algorithm for fast optimization of a path functional of particular form over all possible peak sequences is discussed in the Appendix.

\section{Background and notation}\label{sec:background}

In what follows, we denote by $\hat{f}(\xi)$ and $f^{+}(t)$ the Fourier transform of the function $f(t)$ and its positive frequency part, respectively:
\begin{equation}\label{nt}
\begin{gathered}
\hat{f}(\xi)=\int_{-\infty}^{\infty} f(t)e^{-i\xi t}dt\Leftrightarrow f(t)=\frac{1}{2\pi}\int_{-\infty}^{\infty}\hat{f}(\xi)e^{i\xi t}d\xi,\\
f^{+}(t)\equiv \frac{1}{2\pi}\int_0^\infty \hat{f}(\xi)e^{i\xi t}d\xi,
\end{gathered}
\end{equation}

Next, by an AM/FM component (or simply component) we will mean a signal of the form:
\begin{equation}\label{amfm}
x(t)=A(t)\cos\phi(t)\;\;(\forall t:\;A(t)>0,\;\phi'(t)>0),
\end{equation}
which is additionally required to satisfy $A(t)e^{i\phi(t)}\approx2[A(t)e^{i\phi(t)}]^+$, so that $A(t)$ and $\phi(t)$ are determined uniquely and, in the case of a single component, can be found using the analytic signal approach; a more detailed discussion of issues related to the definition and estimation of the amplitude $A(t)$, phase $\phi(t)$ and instantaneous frequency $\nu(t)\equiv\phi'(t)$ of the component can be found in \cite{Boashash:92,Vakman:96,Picinbono:97,Chen:14,Iatsenko:tfr}.

In real cases, a signal usually contains many components $x_i(t)$ of the form (\ref{amfm}), as well as some noise $\zeta(t)$ (that can be of any form, and is not necessarily white and Gaussian \cite{Chen:14}):
\begin{equation}\label{sig}
s(t)=\sum_i x_i(t)+\zeta(t),
\end{equation}
The goal of ridge analysis is to extract these components, either all or only those of interest, from the signal's TFR.

The two main linear TFRs suitable for components extraction and reconstruction are the windowed Fourier transform (WFT) $G_s(\omega,t)$ and the wavelet transform (WT) $W_s(\omega,t)$. Given a signal $s(t)$, they can be constructed as
\begin{equation}\label{wftwt}
\begin{aligned}
G_s(\omega,t)\equiv&\int_{-\infty}^\infty s^{+}(u)g(u-t)e^{-i\omega (u-t)}du\\
=&\frac{1}{2\pi}\int_0^\infty e^{i\xi t}\hat{s}(\xi)\hat{g}(\omega-\xi)d\xi,\\
W_s(\omega,t)\equiv&\int_{-\infty}^\infty s^{+}(u)\psi^*{\Big(}\frac{\omega(u-t)}{\omega_\psi}{\Big)}\frac{\omega du}{\omega_\psi}\\
=&\frac{1}{2\pi}\int_0^\infty e^{i\xi t}\hat{s}(\xi)\hat{\psi}^*(\omega_{\psi}\xi/\omega)d\xi,
\end{aligned}
\end{equation}
where $s^+(t)$ is the positive frequency part of the signal (as defined in (\ref{nt})), $g(t)$ and $\psi(t)$ are respectively the window and wavelet functions chosen, and $\omega_\psi\equiv\operatorname{argmax}|\hat{\psi}(\xi)|$ denotes the wavelet peak frequency (for the WFT we assume $\operatorname{argmax}|\hat{g}(\xi)|=0$). Note that the WT is commonly defined through the scales $a=\omega_\psi/\omega$, but that in (\ref{wftwt}) we have already transformed to frequencies.

The main difference between the two TFRs mentioned is that the WFT distinguishes the components on the basis of their frequency differences (linear frequency resolution), while the WT does so on the basis of ratios between their frequencies (logarithmic frequency resolution). In effect, while the time-resolution of the WFT is fixed, for the WT it is linearly proportional to frequency, so that the time-modulation of the higher frequency components is represented better than that for the components at lower frequencies.

In numerical simulations we use a Gaussian window for the WFT and a lognormal wavelet for the WT:
\begin{equation}\label{winwav}
\begin{gathered}
\hat{g}(\xi)=e^{-(f_0\xi)^2/2}\Leftrightarrow g(t)=\frac{1}{\sqrt{2\pi}f_0}e^{-(f_0^{-1}t)^2/2},\\
\hat{\psi}(\xi)=e^{-(2\pi f_0\log\xi)^2/2},\quad\omega_\psi=1,\\
\end{gathered}
\end{equation}
where $f_0$ is the resolution parameter determining the tradeoff between time and frequency resolution of the resultant transform (we use $f_0=1$ by default). While the methods developed below are generally applicable for any window/wavelet, the forms (\ref{winwav}) seem to be the best choice \cite{Iatsenko:tfr}, at least for the extraction and reconstruction of components.

\begin{figure*}[t]
\includegraphics[width=1.0\linewidth]{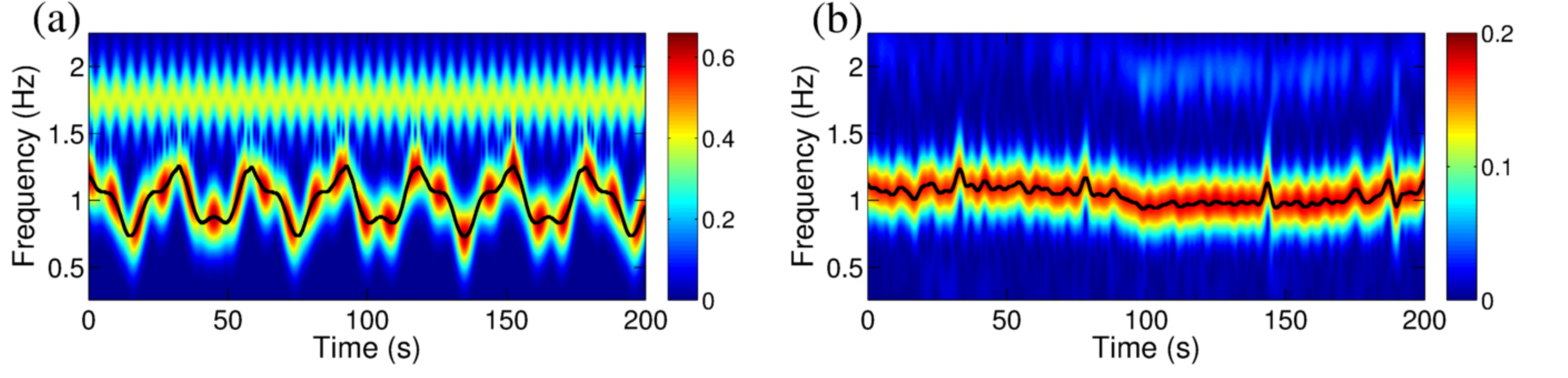}
\caption{Windowed Fourier transforms (WFTs): (a) of the signal $s(t)=\big(1+\frac{1}{3}\cos\frac{2\pi t}{9}\big)\cos\big(2\pi t+6\sin\frac{2\pi t}{30}+\cos\frac{2\pi t}{12})+0.8\cos\big(2\pi \times1.75t+0.5\sin\frac{2\pi t}{5}\big)$; and (b) of the electrocardiogram (ECG) signal. Black lines show the ridge curves $\omega_p(t)$, i.e.\ the sequence of the WFT amplitude peaks, corresponding to the dominant component in each case.}
\label{fig:wftex}
\end{figure*}

As illustrated in Fig.\ \ref{fig:wftex}, the components present in the signal appear in its TFR as ``curves'' (which will be referred to as \emph{ridge curves}), i.e.\ time sequences of close peaks. The problem of curve extraction therefore lies in selecting from among all possible trajectories the sequence of peaks that corresponds to a single component; the positions of these peaks then form a specific frequency profile, which will be denoted as $\omega_p(t)$. Having found the ridge curve, the parameters of the corresponding component can be estimated in a number of ways \cite{Iatsenko:tfr,Lilly:10,Daubechies:11}. In the present work, however, we concentrate on curve extraction only and, except where it is unavoidable, do not consider the reconstruction issues; for a detailed study of the latter, see \cite{Iatsenko:tfr}. Note that, in practice, it is convenient to find the ridge curve associated with the dominant component present, which can then be reconstructed and subtracted from the signal; the procedure can then be repeated to extract any other possible ridge curves.

In what follows, we denote the ridge frequencies, i.e.\ positions of the peaks at each time, as $\nu_m(t)$, the corresponding TFR amplitudes as $Q_m(t)$, and their numbers as $N_p(t)$:
\begin{equation}\label{ppos}
\begin{aligned}
&\nu_m(t):\;
\left\{\begin{array}{l}
\Big[\partial_\omega |H_s(\omega,t)|\Big]_{\omega=\nu_m(t)}=0,\\
\Big[\partial_\omega^2 |H_s(\omega,t)|\Big]_{\omega=\nu_m(t)}<0,\\
\end{array}\right.\\
&Q_m(t)\equiv |H_s(\nu_m(t),t)|,\quad
m=1,...,N_p(t),\\
\end{aligned}
\end{equation}
where $H_s(\omega,t)$ is the chosen TFR of a given signal (WFT $G_s(\omega,t)$ or WT $W_s(\omega,t)$). The ridge curve can then be parametrized as $\omega_p(t)=\nu_{m_c(t)}(t)$, where $m_c(t)$ is the sequence of selected peak indices at each time $t$, which we need to find. Note that the number of peaks $N_p(t)$ can vary in time and in practice is often greater than the number of components present in the signal, with the additional peaks being attributable e.g.\ to noise.

For simplicity, we have treated $\omega$ and $t$ as continuous variables. In practice, however, both time and frequency are discretized, and so also are many other related quantities (e.g.\ the ridge curve $\omega_p(t)$ becomes a discrete set of points). In what follows we therefore assume that the signal is sampled at $t_n=(n-1)\Delta t$ for $n=1,...,N$, so that $N$ is the signal's length in samples, while the TFRs (\ref{wftwt}) are calculated for the frequencies $\omega_k=\omega_{\min}+(k-1)\Delta\omega$ (WFT) or $\omega_k=2^{\frac{k-1}{n_v}}\omega_{\min}$ (WT), where $k=1,...,N_f$. The discretization parameters $\Delta\omega$ and $n_v$ are generally selected by the user, but one can use e.g.\ the criteria suggested in \cite{Iatsenko:tfr} to make an appropriate choice.

\section{Curve extraction schemes}\label{sec:schemes}

The most straightforward way to extract the ridge curve is to first choose some starting point $\omega_p(t_0)$, and then follow from it forward and backward in time, selecting next ridges as those maximizing some suitably chosen functional of the corresponding peak amplitudes and the previously selected ridges. This approach, which we will call \emph{one-step optimization}, can be formulated mathematically as
\begin{equation}\label{onestepopt}
\begin{aligned}
&\mbox{\texttt{for $n=n_0+1,\dots,N$ do:}}\\
&
\begin{aligned}
m_c(t_n)=\underset{m}{\operatorname{argmax}}\Big\{F\big[&t_n,Q_m(t_n),\nu_m(t_n),\\
&\omega_p(t_{n-1}),\omega_p(t_{n-2}),\dots,\omega_p(t_{n_0})\big]\Big\}
\end{aligned}\\
&\omega_p(t_n)=\nu_{m_c(t_n)}(t_n),\\
\end{aligned}
\end{equation}
and similarly backwards in time, for $n=n_0-1,n_0-2,\dots,1$. In (\ref{onestepopt}), $n_0$ denotes the discrete index of the starting time $t_0$ (for which $\omega_p(t_{n_0})$ is known), and the $F\big[...\big]$ is the chosen functional of the current discrete time $t_n$, the peak positions $\nu_m(t_n)$ and amplitudes $Q_m(t_n)$ at this time, and all previously selected ridge points $\{\omega_p(t_{n_0}\leq t\leq t_{n-1})\}$. For scheme (\ref{onestepopt}) to be O$(N)$, the functional $F\big[...\big]$ should either depend on the finite number of previously selected ridges, or on the set of parameters which can be updated in O$(1)$ steps whenever new point becomes available (e.g.\ the moments of $\omega_p(t)$).

To implement (\ref{onestepopt}), one needs to choose the starting time index $n_0$ and the corresponding ridge $\omega_p(t_{n_0})$. It seems natural to select this starting point (among all times and ridges) as being that for which the functional in (\ref{onestepopt}) is likely to attain its maximum:
\begin{equation}\label{spoint}
\begin{aligned}
\omega_p(t_{n_0})=&\nu_{m_0}(t_{n_0}),\\
\{m_0,n_0\}=&\underset{\{m,n\}}{\operatorname{argmax}}\Big\{F_0[t_n,Q_m(t_n),\nu_m(t_n)]\Big\},
\end{aligned}
\end{equation}
where $F_0[...]$ denotes a ``zero-step'' version of the original functional, obtained from the latter by taking its maximum among all the other parameters. For example, if one has $F[...]=f(Q_m(t_n),\nu_m(t_n))+g(\nu_m(t_n)-\omega_p(t_{n-1}))$, then $F_0[...]=f(Q_m(t_n),\nu_m(t_n))+\max_{\Delta\xi} g(\Delta\xi)$; if additionally $f(Q_m(t_n),\nu_m(t_n))$ does not depend on $\nu_m(t_n)$ and is proportional to $Q_m(t_n)$, then (\ref{spoint}) will correspond to the highest TFR amplitude peak over all times. The criterion (\ref{spoint}) works well in most cases, although it could still provide a ``bad'' starting point when the sharp time events are present or the noise is too strong.

A serious drawback of the outlined one-step approach (\ref{onestepopt}) is that even a single wrongly selected point might completely change all the following curve being extracted. Consequently, it is more accurate to optimize the functional not over each consecutive point, as in (\ref{onestepopt}), but over the whole profile $\omega_p(t)$, selecting the ridge curve as that which maximizes the full integral of $F[...]$ over time:
\begin{equation}\label{pathopt}
\begin{aligned}
\{\omega_p(t_1),...,\omega_p(t_N)\}=&\{\nu_{m_c(t_1)}(t_1),...,\nu_{m_c(t_N)}(t_N)\},\\
\{m_c(t_1),...,m_c(t_N)\}=&\underset{\{m_1,m_2,...,m_N\}}{\operatorname{argmax}}\sum_{n=1}^N F\big[t_n,Q_{m_n}(t_n),\\
&\nu_{m_n}(t_n),\big\{\nu_{m_1}(t_1),...,\nu_{m_N}(t_N)\big\}\big].\\
\end{aligned}
\end{equation}
This approach, where the optimization is performed over all possible sequences of peak numbers $\{m_1,m_2,...,m_N\}$, will be referred to as the \emph{path optimization}. In general, it is computationally very expensive. However, if the functional depends on only a finite number of previous points $\{\omega_p(t_{n-i}),...,\omega_p(t_{n-1})\}$ rather than the full history, then the optimal path in terms of (\ref{pathopt}) can be selected in O$(N)$ computations using dynamic programming algorithm (see Appendix). Note that, in this way, the widely-used method of Carmona et.\ al.\ \cite{Carmona:97} can also be performed in O$(N)$ steps instead of using the computationally-expensive simulated annealing, as previously.

As will be seen, the path optimization (\ref{pathopt}) usually gives much better results than the one-step optimization (\ref{onestepopt}), and should therefore always be preferred to the latter. Furthermore, it has no problem associated with the selection of the starting point (\ref{spoint}), as all the trajectories are explored.

What remains is to select an appropriate functional in (\ref{pathopt}). We consider below some curve-extraction schemes defined by particular classes of $F[...]$. We first develop these schemes for the WFT, and then discuss how they can be adjusted for the WT. In all cases, we perform path optimization using the algorithm discussed in the Appendix. Taking into account its complexity, and the fact that one needs to locate all peaks (\ref{ppos}) in the TFR prior to applying any extraction procedure, the computational costs of the methods discussed below are O$(N_f N)$+O$(M_p^2 N)$ (scheme I) and O$(N_f N)$+O$(M_p^2 N \log N)$ (scheme II), with $\log N$ corresponding to the number of iterations as discussed below; $N_f$ and $M_p\equiv \max_t N_p(t)$ are respectively the number of frequencies $\omega_k$ for which the TFR is calculated, and the maximum number of TFR amplitude peaks present at any one time. Both $N_f$ and $M_p$ are independent of $N$.

\begin{remark}\label{rem:deff}
Because in practice the frequency scale for the WFT/WT is discretized, the ridge frequencies $\nu_m(t)$ also take discrete values at each time. As a result, e.g.\ the differences between consecutive ridges $\Delta\omega_p(t_n)\equiv\omega_p(t_n)-\omega_p(t_{n-1})$ cannot reliably be calculated, being ``quantized'' in steps determined by the widths of the frequency bins. To avoid consequential problems, we use parabolic interpolation (based on the TFR amplitude at the corresponding peak and in the two adjacent bins) to find peak positions $\nu_m(t)$ more precisely. Because the TFR amplitudes take continuous values, the estimates of $\nu_m(t)$ (and therefore those of $\Delta\omega_p(t)$) also become continuous. One then does not need to worry about the related discretization effects, which could otherwise influence significantly the performance of methods that are based on the differences between ridge frequencies.
\end{remark}

\subsection{Scheme I($\alpha$): penalization of frequency jumps}\label{sec:scheme1}

A widespread approach is to penalize the frequency difference between the consecutive ridge points, so that
\begin{equation}\label{es1}
F\big[...\big]=\log Q_m(t_n)+w(\nu_m(t_n)-\omega_p(t_{n-1}),\alpha),
\end{equation}
where $w(\Delta\xi,\alpha)$ is some weighting function, aimed at suppressing frequency jumps, and $\alpha$ is its set of adjustable parameters. Note that in (\ref{es1}) one can choose another function of $Q_m(t_n)$ instead of the logarithm, e.g.\ $|Q_m(t_n)|^2$; however, the logarithm seems to be the most appropriate because, in this case, the path functional (\ref{pathopt}) depends on the product of all amplitudes and thus can be significantly influenced even by a single ``wrong'' point, making selection of the latter less probable.

The class of functionals (\ref{es1}) is a very popular choice. Thus, the approach of \cite{Barros:01} corresponds to $w(\Delta\xi,\alpha)=0$ for $\Delta\xi\in[-1/\alpha,1/\alpha]$ and $=-\infty$ otherwise, while the procedure used in \cite{Thakur:13,Chen:14} utilizes the quadratic weights $w(\Delta\xi,\alpha)=\alpha\Delta\xi^2$ (though in these methods the optimization is carried out over all frequency bins $\omega_k$ at each time rather than using only the peaks $\nu_m(t)$, as we do here). The algorithm of \cite{Iatsenko:cardio} also represents a variant of (\ref{es1}). Finally, the approach of Carmona et. al. \cite{Carmona:97} can be viewed as a modified version of (\ref{es1}) with additional penalization of the second order frequency differences.

The main disadvantage of the approaches mentioned is that they require fine tuning of each method's parameters to obtain an accurate result, with different choices being needed for different signals and different characteristics of the TFR in use. To make the parametrization more universal, the weighting function should utilize the resolution properties of the WFT, which are determined by the window function $g(t)$. Thus, for a given window there exists a minimum frequency (resp. time) difference $\Delta\xi_g$ (resp. $\Delta\tau_g$) for which two frequency events, e.g.\ tones (resp. time events, e.g.\ delta-peaks) can be resolved in the WFT. In other words, the larger $\Delta\tau_g$ (smaller $f_0$ in (\ref{winwav})) is, the less time-variability is allowed for the components (so one expects smaller frequency jumps).

We therefore penalize the ratio of the observed time-derivative of the ridge frequency difference to its characteristic value, which can naturally be taken as $\Delta\xi_g/\Delta\tau_g$. This leads to the choice
\begin{equation}\label{wf1}
w(\Delta\xi,\alpha)=\alpha\widetilde{w}\Big(\frac{f_s|\Delta\xi|}{\Delta\xi_g/\Delta\tau_g}\Big)
=-\alpha\frac{f_s|\Delta\xi|}{\Delta\xi_g/\Delta\tau_g},
\end{equation}
where $f_s$ is the signal sampling frequency, while $\Delta\xi_g$ and $\Delta\tau_g$ are chosen resolution measures. We use those introduced in \cite{Iatsenko:tfr}, taking $\Delta\xi_g$ and $\Delta\tau_g$ as the widths of the regions in time and frequency encompassing 50\% of the window function:
\begin{equation}\label{WFTres}
\begin{gathered}
\Delta\xi_g=\xi_g^{(2)}(0.5)-\xi_g^{(1)}(0.5),\;\;\;\Delta\tau_g=\tau_g^{(2)}(0.5)-\tau_g^{(1)}(0.5);\\
\xi_g^{(1,2)}(\epsilon):\; |R_g(\xi\leq\xi_g^{(1)})|<\epsilon/2,\;|1-R_g(\xi\geq\xi_g^{(2)})|<\epsilon/2;\\
\tau_g^{(1,2)}(\epsilon):\; |P_g(\tau\leq\tau_g^{(1)})|<\epsilon/2,\;|1-P_g(\tau\geq\tau_g^{(2)})|<\epsilon/2;\\
R_g(\omega)\equiv\frac{\int_{-\infty}^\omega\hat{g}(\xi)d\xi}{\int_{-\infty}^\infty\hat{g}(\xi)d\xi},\quad
P_g(\tau)\equiv\frac{\int_{-\infty}^\tau g(t)dt}{\int_{-\infty}^\infty g(t)dt}.
\end{gathered}
\end{equation}
For a Gaussian window (\ref{winwav}) one obtains $\Delta\xi_g/\Delta\tau_g=1/f_0^2$, and this result remains the same even if using as $\Delta\xi_g$ and $\Delta\tau_g$ the conventional standard deviations of $|\hat{g}(\xi)|^2$ and $|g(t)|^2$, respectively.

With the choice (\ref{wf1}), the parameter $\alpha$ is expected to be relatively universal, so that the same value should work well for different window functions. Note that, although in (\ref{wf1}) we use $\widetilde{w}(r)=-|r|$, other functions can be utilized instead. However, for any reasonable choice, the method remains qualitatively the same, i.e.\ one expects it to suffer from the same drawbacks and to have similar issues.

It is important to note, that scheme I corresponds to simple cases of ``global maximum'' and ``nearest neighbour'' curve extraction for $\alpha=0$ and $\alpha\rightarrow\infty$, respectively:
\begin{itemize}
\item \emph{Global Maximum ($\alpha=0$).} In this case the functional (\ref{es1}) reduces to $F[...]=\log Q_m(t_n)$, so that the maximum peak will be selected at each time, taking no account of the previous ridge points.
\item \emph{Nearest Neighbour ($\alpha\rightarrow\infty$).} This case differs for one-step optimization (\ref{onestepopt}) and path optimization (\ref{pathopt}). The former approach corresponds to selecting at each new step the peak which is nearest to the previous one, taking no account of its amplitude. The latter approach will give simply the least frequency-varying curve (which is a rather pathological case).
\end{itemize}

\subsection{Scheme II($\alpha$,$\beta$): adaptive parametrization}\label{sec:scheme2}

In the previous scheme, there is an adjustable parameter $\alpha$ that determines the suppression of the frequency variations. Although some choices (e.g.\ $\alpha=1$) appear to be relatively universal, they still remain highly non-adaptive, so that a particular parameter value might be suitable for one type of the signal, and a different value for another type. For example, in the case of chirps $\sim \cos(at+bt^2)$ it is clear that one should penalize not the frequency jumps, but their differences from the actual frequency growth rate.

To make the scheme adaptive, the parameters of the functional should be matched to the properties of the component being extracted, such as the typical variations of its instantaneous frequency. The latter can be characterized by the averages and standard deviations of the ridge frequencies $\omega_p(t)$ and their differences $\Delta\omega_p(t_n)\equiv\omega_p(t_n)-\omega_p(t_{n-1})$; or, which appears to be more stable in practice, by the corresponding medians $\mathfrak{m}[...]$ and 50\% ranges $\mathfrak{s}[...]$, defined for an arbitrary function $f(t)$ as
\begin{equation}\label{medrange}
\mathfrak{m}[f(t)]\equiv\underset{0.5}{\rm perc}[f(t)],\quad
\mathfrak{s}[f(t)]\equiv\underset{0.75}{\rm perc}[f(t)]-\underset{0.25}{\rm perc}[f(t)],
\end{equation}
where ${\rm perc}_p[f(t)]$ denotes the $p^{\rm th}$ quantile of $f(t)$.

An adaptive functional can then be constructed by suppressing not the absolute frequency jumps, as before, but the relative deviations of the component's ridge frequency and its derivative from their typical values:
\begin{equation}\label{es2}
\begin{aligned}
F\big[...\big]=&\log Q_m(t_n)+w_2\big(\nu_m(t_n),\mathfrak{m}[\omega_p],\mathfrak{s}[\omega_p],\beta\big)\\
&+w_1\big(\nu_m(t_n)-\omega_p(t_{n-1}),\mathfrak{m}[\Delta\omega_p],\mathfrak{s}[\Delta\omega_p],\alpha\big).\\
\end{aligned}
\end{equation}
As previously, we choose the first order penalization functions
\begin{equation}\label{wf2}
\begin{gathered}
w_1\big(\Delta\xi,\mathfrak{m}[\Delta\omega_p],\mathfrak{s}[\Delta\omega_p],\alpha\big)
=-\alpha\Big|\frac{\Delta\xi-\mathfrak{m}[\Delta\omega_p]}{\mathfrak{s}[\Delta\omega_p]}\Big|,\\
w_2\big(\xi,\mathfrak{m}[\omega_p],\mathfrak{s}[\omega_p],\beta\big)
=-\beta\Big|\frac{\xi-\mathfrak{m}[\omega_p]}{\mathfrak{s}[\omega_p]}\Big|.\\
\end{gathered}
\end{equation}
By maximizing the path integral (\ref{pathopt}) based on the functional (\ref{es2}), one is in fact trying to extract the curve which is most consistent with itself. Thus, the strength of the respective frequency variations becomes not important, and it is only their agreement and similarity at different times that matters.

Even the most adaptive method can be parametrized to tackle special cases, and in (\ref{es2}) we have introduced the adjustable parameters $\alpha$ and $\beta$ controlling the strengths of suppression of the corresponding relative deviations. However, although there are now two parameters, they are in fact more universal than the single parameter of scheme I. Thus, the particular choice of $\alpha,\beta$ for scheme II is expected to work well for a larger class of signals than the particular choice of $\alpha$ in the scheme I, as will be seen below. This is because in (\ref{es2}) we take explicitly into account the actual properties of the component being extracted, penalizing deviations from its typical behavior rather than simply the frequency jumps. Additionally, by suppressing the relative deviations of the component's frequency from its mean, scheme II stabilizes the curve in its characteristic frequency range (thus lowering the possibility that it will ``escape'' and switch to another component), while there is no such mechanism in scheme I.

The functional (\ref{es2}) depends, however, on the whole time-evolution of $\omega_p(t)$, so that the path optimization (\ref{pathopt}) cannot be performed in O$(N)$ steps, as before (see Appendix); nor is it evident how to update the functional at each step if using the one-step optimization (\ref{onestepopt}). Nevertheless, one can approach the approximately optimal curve $\omega_p(t)$ by use of a kind of fixed point iteration \cite{Agarwal:01}. Thus, starting with some initial guess $\omega_p^{(0)}(t)$, one calculates the corresponding medians and ranges, fixes them in (\ref{es2}) (so that the functional now depends on only two consecutive ridges rather than on the full history, meaning that the algorithm discussed in Appendix becomes applicable), and extracts the newer profile $\omega_p^{(1)}(t)$ in the usual way. The (fixed) medians and ranges are then updated to those of the $\omega_p^{(1)}(t)$ and, based on these newer estimates, the next approximation $\omega_p^{(2)}(t)$ is found in the same manner. The procedure is repeated until the curves obtained in two consecutive iterations coincide perfectly ($\omega_p^{(n)}(t)=\omega_p^{(n-1)}(t)$ for all $t$). For the first iteration, we use a simple Global Maximum curve $\omega_p^{(0)}(t)={\rm argmax}_\omega|G_s(\omega,t)|$.

The convergence of the fixed-point algorithm outlined above is in general hard to prove. In practice, however, the procedure converges not only exactly (so that the next iterations produce absolutely identical curves), but also rapidly. To show this, we have analysed the performance of the method for white noise signals with different sampling frequencies and time lengths, thus trying to model the worst case (as the method will obviously converge faster if the signal contains some pronounced components). The results are presented in Supplementary Material (Fig.\ 1 there). The number of iterations needed is always relatively small, being proportional to $\log N$; it is determined primarily by the signal's time length, while the sampling frequency only has a very minor effect. Note also that one can set some maximum number of allowed iterations if desired, though in our simulations the procedure always converged exactly and rapidly.


\subsection{Adjustments for the WT}\label{sec:adjustwt}

Due to the logarithmic frequency resolution of the WT, one should consider not the frequencies but their logarithms, which is the only significant difference from the WFT case. Thus, in the case of the WT one uses the same schemes and functionals, but now everything is taken on a logarithmic frequency scale ($\omega_p(t_n)\rightarrow\log\omega_p(t_n)$, $\Delta\omega_p(t_n)\equiv\omega_p(t_n)-\omega_p(t_{n-1})\rightarrow\Delta\log\omega_p(t_n)\equiv\log\omega_p(t_n)-\log\omega_p(t_{n-1})$, and similarly for all the other frequency variables). We now summarize briefly the required adjustments.

\textbf{Scheme I:} Instead of $w(\nu_m(t_n)-\omega_p(t_{n-1}),\alpha)$, in (\ref{es1}) one uses $w(\log\nu_m(t_n)-\log\omega_p(t_{n-1}),\alpha)$. The form of the penalization function (\ref{wf1}) remains qualitatively the same:
\begin{equation}
w(\Delta\log\xi,\alpha)=\alpha\widetilde{w}\Big(\frac{f_s|\Delta\log\xi|}{\Delta\log\xi_\psi/\Delta\tau_\psi}\Big)
=-\alpha\frac{f_s|\Delta\log\xi|}{\Delta\log\xi_\psi/\Delta\tau_\psi},
\end{equation}
but one now uses the wavelet's characteristic log-frequency and time differences $\Delta\log\xi_\psi$ and $\Delta\tau_\psi$, respectively. We use the estimates given in \cite{Iatsenko:tfr}, which are calculated as
\begin{equation}\label{WTres}
\begin{gathered}
\Delta\log\xi_\psi=\log\frac{\xi_\psi^{(2)}(0.5)}{\xi_\psi^{(1)}(0.5)},\;\;\;\Delta\tau_\psi=\tau_\psi^{(2)}(0.5)-\tau_\psi^{(1)}(0.5);\\
\xi_\psi^{(1,2)}(\epsilon):\; |R_\psi(\xi\leq\xi_\psi^{(1)})|<\epsilon/2,\;|1-R_\psi(\xi\geq\xi_\psi^{(2)})|<\epsilon/2;\\
\tau_\psi^{(1,2)}(\epsilon):\; |P_\psi(\tau\leq\tau_\psi^{(1)})|<\epsilon/2,\;|1-P_\psi(\tau\geq\tau_\psi^{(2)})|<\epsilon/2;\\
R_\psi(\omega)\equiv\frac{\int_0^\omega\hat{\psi}^*(\xi)d\xi/\xi}{\int_0^\infty\hat{\psi}^*(\xi)d\xi/\xi},\quad
P_\psi(\tau)\equiv\frac{\int_{-\infty}^\tau \psi^*(t)e^{i\omega_\psi t}dt}{\int_{-\infty}^\infty \psi^*(t)e^{i\omega_\psi t}dt}.
\end{gathered}
\end{equation}

\textbf{Scheme II:} In (\ref{es2}) the $w_1(...)$ and $w_2(...)$ are changed to $w_1(\log\nu_m(t_n)-\log\omega_p(t_{n-1}),\mathfrak{m}[\Delta\log\omega_p],\mathfrak{s}[\Delta\log\omega_p])$ and $w_2(\log\nu_m(t_n),\mathfrak{m}[\log\omega_p],\mathfrak{s}[\log\omega_p])$, respectively, with their basic forms (\ref{wf2}) remaining the same.

\section{Comparison of schemes}\label{sec:comparison}

\subsection{Test signals}

\begin{figure*}[t]
\includegraphics[width=1.0\linewidth]{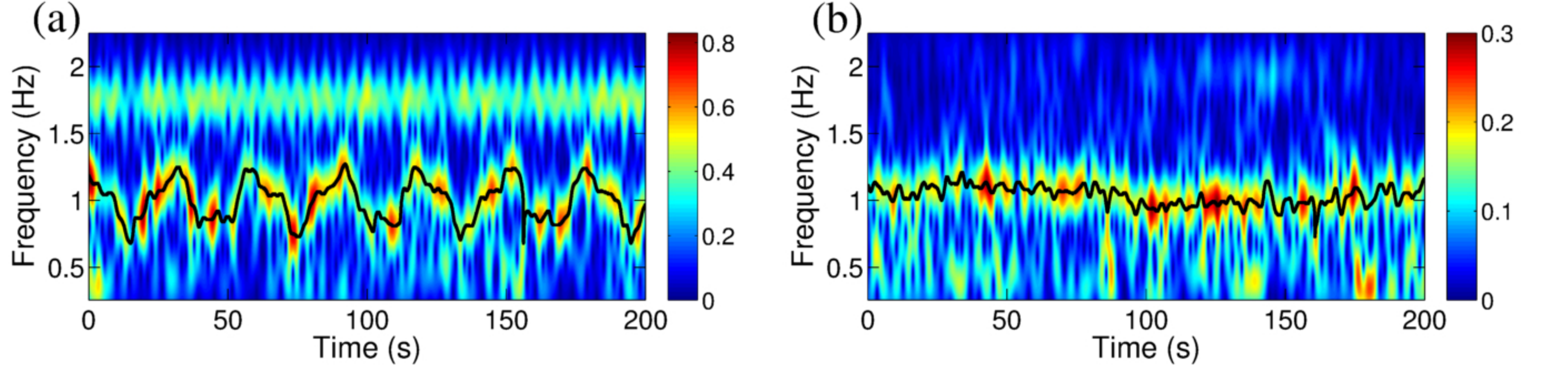}
\caption{WFTs of the same signals as in Fig.\ \ref{fig:wftex}, but additionally corrupted by noise of the form (\ref{signoise}), with a standard deviation $\sigma=0.6$ for the signal corresponding to (a), and $\sigma=0.3$ for the signal corresponding to (b).}
\label{fig:wftexnoise}
\end{figure*}

We now test the relative performances of the different methods on two signals. The first signal is an AM/FM component with simple sinusoidal amplitude modulation and two-sinusoidal frequency modulation, plus a weaker component:
\begin{equation}\label{s1}
\begin{aligned}
s_1(t)=&\big(1+\frac{1}{3}\cos\frac{2\pi t}{9}\big)\cos\big(2\pi t+6\sin\frac{2\pi t}{30}+\cos\frac{2\pi t}{12})\\
&+0.8\cos\big(2\pi\times 1.75t+0.5\sin\frac{2\pi t}{5}\big).
\end{aligned}
\end{equation}
Note that, although an AM/FM component around 1 Hz is dominant in terms of both maximum amplitude and mean squared amplitude, there are certain times at which the amplitude of the other component (at around 1.75 Hz) becomes higher, thereby introducing additional complications for the curve extraction. The second test signal is taken from real life, representing the central 200 s part of a 30 min electrocardiogram (ECG) signal recorded from a 30 years old male subject \cite{Iatsenko:cardio}. The WFTs for both signals are shown above in Fig.\ \ref{fig:wftex}.

The main complications that arise in curve extraction relate to the appearance of other WFT amplitude peaks near $\omega_p(t)$, which can be due either to noise or to other components. We model these complications by corrupting the signal with colored noise $\eta(t)$ of unit deviation and a particular Fourier amplitude (while the phases of its Fourier coefficients are random):
\begin{equation}\label{signoise}
s(t)=s(t)+\sigma\eta(t),\quad |\hat{\eta}(\xi)|\sim\frac{1}{4\pi^2+\xi^2}.
\end{equation}
Being asymmetric, the noise amplitude at frequency $0.5$ Hz is around 2.5 times higher than at $1.5$ Hz, corrupting the dominant components (which have a mean frequency around $1$ Hz in both test signals) unequally in frequency on the two sides. This gives an opportunity to study reliably the relative performance of the different methods, as colored noise can additionally model the effect of other components that are asymmetrically distributed in frequency around the component of interest. The WFTs of the two test signals corrupted with noise are presented in Fig.\ \ref{fig:wftexnoise}.

It is well known that, even in the absence of noise, the ridge points are not located exactly at the true instantaneous frequencies $\nu(t)\equiv\phi'(t)$ \cite{Lilly:10,Iatsenko:tfr}. Thus, if we compare the $\omega_p(t)$ obtained with the true frequency profile then, even in the case when the curve extraction works perfectly (e.g.\ when there is a single peak at each time, and hence only one possible ridge curve) there will be some discrepancy between the two. At the same time, what we want to test is how well the methods presented can identify the peak sequence corresponding to the component of interest, and not how well one can then reconstruct the component's parameters from this sequence. Therefore, to assess the performance of the curve identification method, rather than the performance of the TFR itself or the accuracy with which frequencies are estimated from ridges, we compare the extracted $\omega_p(t)$ with the ``ideal'' ridge curve $\widetilde{\omega}_p(t)$ obtained in the noise-free case. The corresponding error $\epsilon_f$ can then be defined as
\begin{equation}\label{ferr}
\epsilon_f^2\equiv\frac{\langle[\omega_p(t)-\widetilde{\omega}_p(t)]^2\rangle}
{\langle[\widetilde{\omega}_p-\langle\widetilde{\omega}_p\rangle]^2\rangle},
\end{equation}
where $\langle...\rangle$ denotes the time-average. An additional complication is that, because noise changes the ridge profile as it appears in the WFT, there always exists some deviation between the extracted profiles with and without noise, which is unrelated to performance of extraction method. Thus, the $\epsilon_f$ (\ref{ferr}) contains both an irreducible, inherent, error related to the effect of noise on the TFR, and the error of the curve extraction method. Therefore, we only compare the performance of different methods, without aiming to find the profile as it would be without noise (which is generally impossible). 

In the simulations, both test signals are sampled at 20 Hz. We will test curve extraction only for the WFT, but the results remain qualitatively the same for the WT as well. To eliminate boundary distortions in the TFR, we simulate the first test signal (\ref{s1}) for 1000 s, calculate the corresponding WFT and then use only its central 200 s part; the same procedure is applied for the ECG signal. We use a Gaussian window (\ref{winwav}) with $f_0=1$ and calculate the WFTs at frequencies $\omega_k/2\pi=0.25+(k-1)\Delta\omega/2\pi\in[0.25,2.25]$ (this range of frequencies is chosen based on {\it a priori} knowledge that all components of interest are contained in it) with $\Delta\omega=\Delta\xi_g/25\approx 2\pi\times0.008$. For both signals, we use 40 noise realizations, which are the same for each method, parameters and noise intensities $\sigma$ being tested.

\begin{remark}
Note that, for the first test signal (\ref{s1}), if one extracts $\omega_p(t)$ corresponding to the weaker component $0.8\cos\big(2\pi \times1.75t+0.5\sin\frac{2\pi t}{5}\big)$, this can also be regarded as a not-bad result. However, we are mainly interested in testing the accuracy with which the parameters of the dominant component (around 1 Hz) can be recovered. Therefore, if the ridge profile $\omega_p(t)$ extracted from the WFT of the first test signal lies closer to the frequency of the non-dominant component, we discard the corresponding peaks and re-extract the curve. This does not apply for the second test signal.
\end{remark}

\begin{figure*}[t!]
\includegraphics[width=1.0\linewidth]{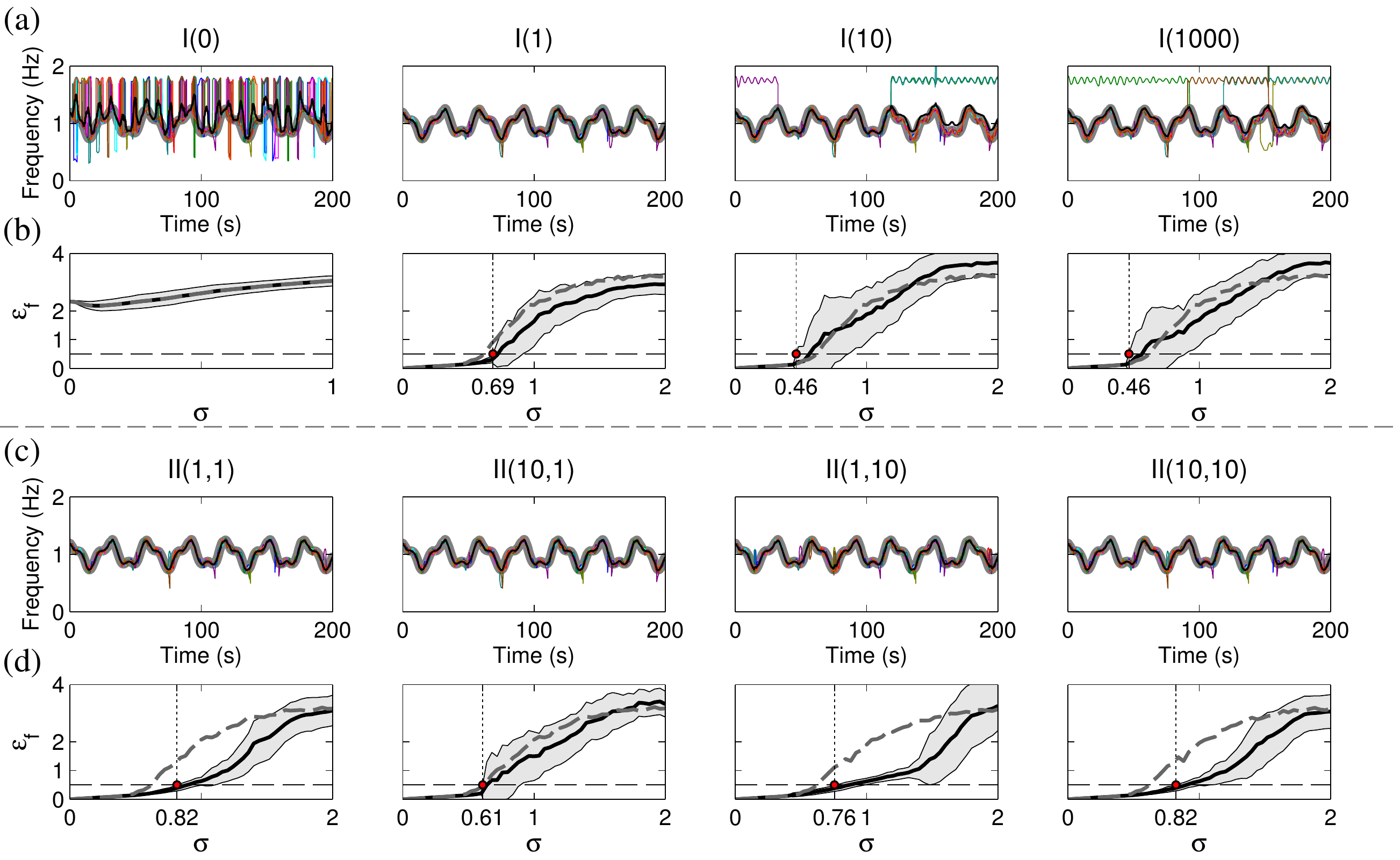}
\caption{Performance of the different schemes for ridge curve extraction from the WFT of the first test signal (\ref{s1}), as illustrated by: (a,c) examples of the extracted $\omega_p(t)$ when the noise standard deviation is $\sigma=0.6$ (the WFT of the particular signal realization at this noise level is presented in Fig.\ \ref{fig:wftexnoise}(a)); (b,d) dependence of the relative error $\epsilon_f$ (\ref{ferr}) on the standard deviation $\sigma$ of the noise. In (a,c) the wide gray background line shows the extracted frequency profile in the noise-free case, the bold black lines correspond to the mean $\omega_p(t)$ over all noise realizations, while the (mostly almost coincident) thin lines show individual extracted curves for 10 (out of 40) noise realizations. In (b,d) the bold black lines show the mean $\epsilon_f$ over all noise realizations, with the gray regions around them indicating $\pm 1$ standard deviation; the bold gray dashed lines show the ensemble mean of $\epsilon_f$ if the schemes were performed using the one-step optimization (\ref{onestepopt}) instead of the (default) path optimization (\ref{pathopt}); vertical dotted lines indicate the values of $\sigma$ for which the mean error plus its standard deviation over noise realizations crosses the level $\epsilon_f=0.5$, shown by horizontal dashed lines.}
\label{fig:s1wft}
\end{figure*}

\subsection{Results}

Results of application of the different curve extraction schemes to the WFT of the first test signal (\ref{s1}) are presented in Fig.\ \ref{fig:s1wft}. The performance of each method is quantified by its maximum tolerable noise level $\sigma_{\max}$, indicated by vertical dotted lines in Fig.\ \ref{fig:s1wft}: we define it as the noise intensity $\sigma$ at which the mean error $\epsilon_f$ (\ref{ferr}) plus its standard deviation over noise realizations reaches 0.5, implying that in many cases the resultant $\omega_p(t)$ is inaccurate. Note that, in each case, the default path optimization (\ref{pathopt}) approach has clear and significant advantages over the one-step optimization (\ref{onestepopt}), with the mean errors for the latter being shown by dashed gray lines in Fig.\ \ref{fig:s1wft}(b,d).

From Fig.\ \ref{fig:s1wft}, it can be seen that the worst performance in the case of the first test signal (\ref{s1}) is exhibited by the I(0) (Global Maximum) method, which is to be expected, given that the amplitude of the weaker component is sometimes higher than that of the dominant one. With increasing $\alpha$ above zero, the performance of the method I($\alpha$) greatly improves (Fig.\ \ref{fig:s1wft}(a,b)), reaching its optimum at some $0<\alpha<10$, and then deteriorating again. Thus, for scheme I and the parameters tested, the best results are achieved at $\alpha=1$.

Nevertheless, much better performance is demonstrated by schemes II(1,1) and II(10,10), which can trace the ridge curve reliably even in the presence of very strong noise. Methods II(10,1) and II(1,10) do not work so well, indicating that large asymmetries between $\alpha$ and $\beta$ are not advantageous, which is to be expected given that we use relative (i.e.\ normalized) deviations.

\begin{figure*}[t!]
\includegraphics[width=1.0\linewidth]{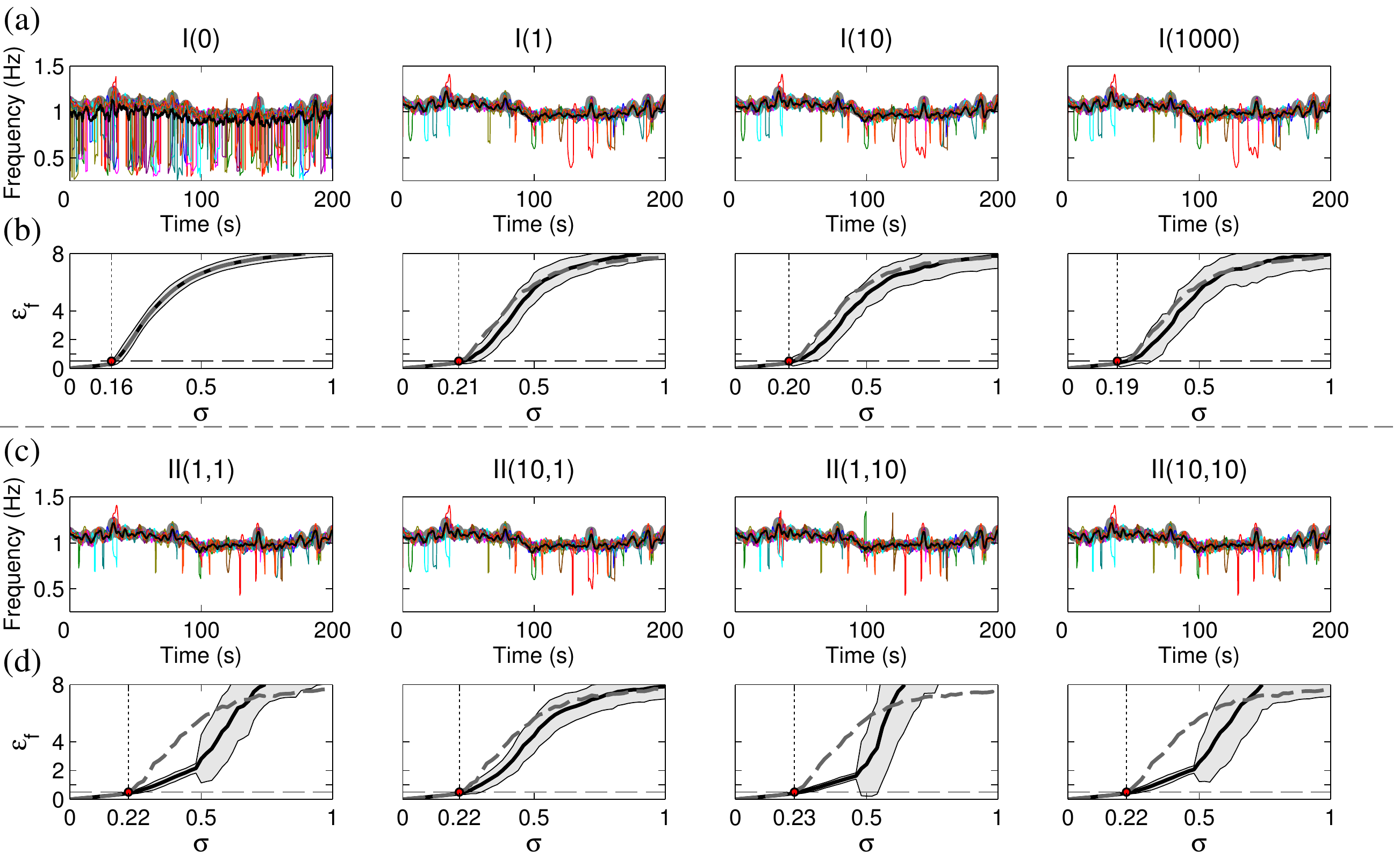}
\caption{Results for the second test signal; otherwise same as Fig.\ \ref{fig:s1wft}. In (a,c) the examples of extracted $\omega_p(t)$ are now shown for $\sigma=0.3$ (the WFT of one particular signal realization at this noise level is presented in Fig.\ \ref{fig:wftexnoise}(b))}.
\label{fig:s2wft}
\end{figure*}

Results for the second test signal, the ECG, are presented in Fig.\ \ref{fig:s2wft}. Clearly, the situation there is similar to the one observed for the first test signal in Fig.\ \ref{fig:s1wft}. However, now the performance of methods I and II is almost independent of their parameters (except I(0)), at least for the parameter values considered.

Summarizing, the best results were achieved with scheme II, in particular II(1,1) and II(10,10). Scheme I($\alpha$) seem to be most accurate for $\alpha=1$ (at least for the parameters tested), while the Global Maximum method, corresponding to I(0), is largely useless and should not be used. In all cases, the path optimization (\ref{pathopt}) approach was superior to the one-step optimization (\ref{onestepopt}).

\section{Extraction of curves from the synchrosqueezed transforms}\label{sec:ss}

\begin{figure*}[t]
\includegraphics[width=1.0\linewidth]{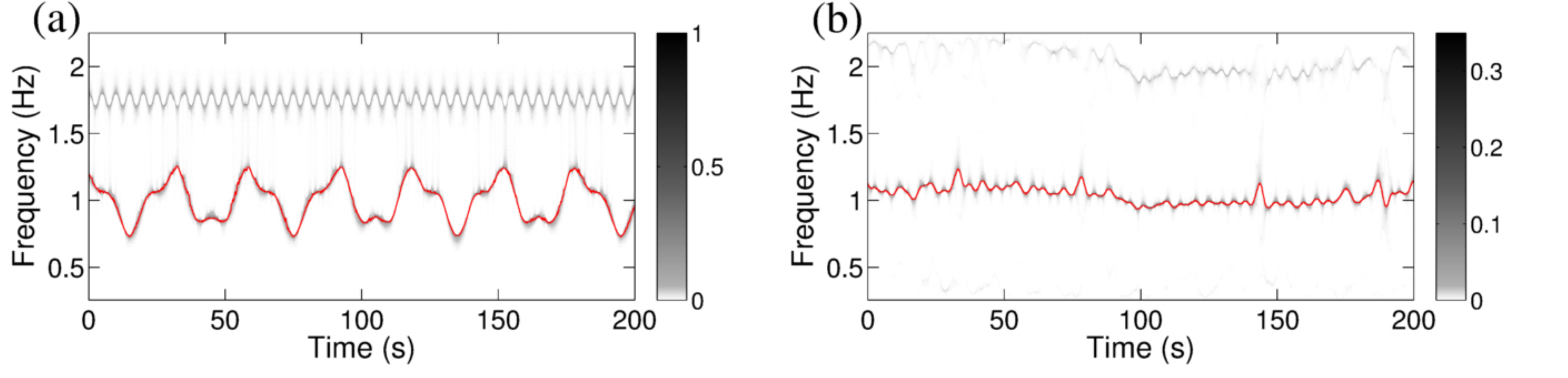}\\
\includegraphics[width=1.0\linewidth]{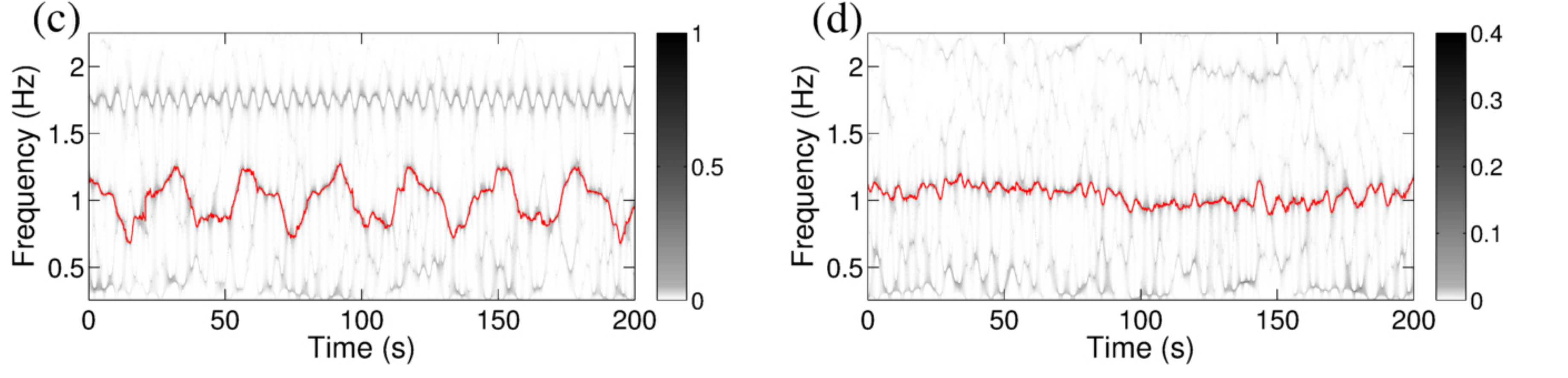}
\caption{Synchrosqueezed WFTs: (a,b) constructed from the WFTs shown in Figs.\ \ref{fig:wftex}; (c,d) constructed from the WFTs shown in Fig.\ \ref{fig:wftexnoise}. Thin red lines show the ridge curves corresponding to the dominant components in each case.}
\label{fig:sswftex}
\end{figure*}

\begin{figure*}[t]
\includegraphics[width=1.0\linewidth]{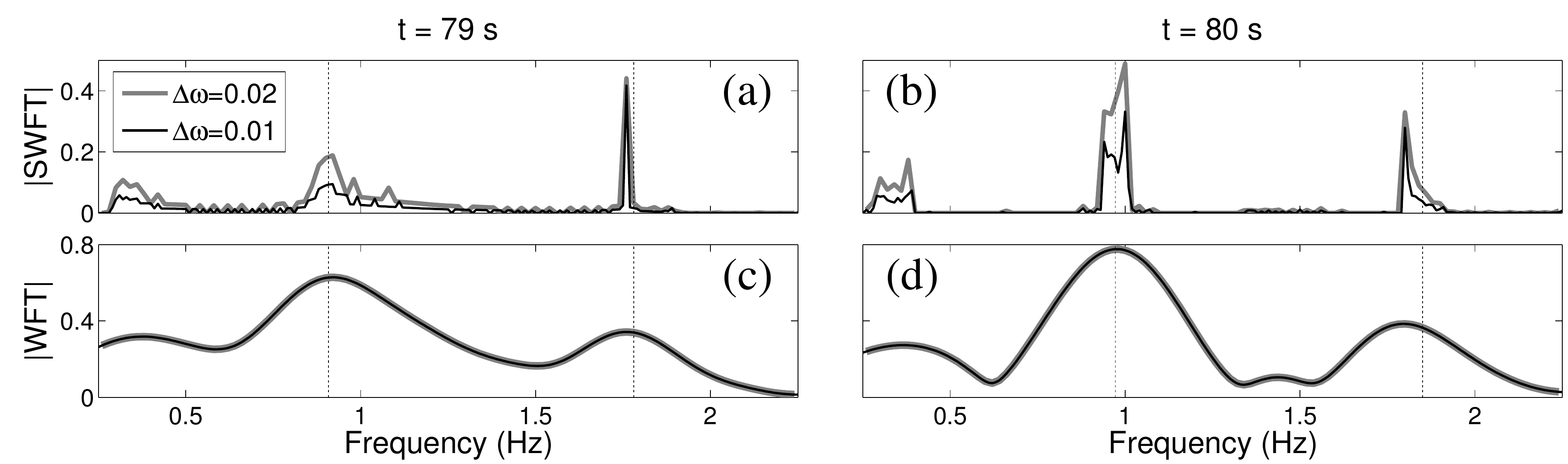}
\caption{Snapshots (a,b) of the SWFT amplitudes and (c,d) of the WFT amplitudes for the first test signal (\ref{s1}) corrupted with noise (\ref{signoise}) of standard deviation $\sigma=0.6$. Thick gray lines and thin black lines show the values obtained using frequency bin widths of $\Delta\omega/2\pi=0.02$ and $\Delta\omega/2\pi=0.01$, respectively. Dotted vertical lines indicate the instantaneous frequencies of each of the two AM/FM components in signal (\ref{s1}) at the corresponding times. This figure shows that, while the WFT peaks (c,d) are generally proportional to the amplitudes of the components, peaks in the SWFT (a,b) depend on the choice of the discretization step $\Delta\omega$ in a nonuniversal and quite sophisticated way.}
\label{fig:wftsswft}
\end{figure*}

Synchrosqueezing \cite{Daubechies:11,Thakur:11,Thakur:13} represents a particular reassignment method \cite{Auger:95,Auger:13} that can be used to construct a more concentrated representation from the WFT and WT by utilizing relationships between the rates of phase growth of the corresponding coefficients. The synchrosqueezed WFT (SWFT) $V_s(\omega,t)$ and synchrosqueezed WT (SWT) $T_s(\omega,t)$ can be constructed as
\begin{equation}\label{ss}
\begin{gathered}
\begin{aligned}
&V_s(\omega,t)=C_g^{-1}\int_{-\infty}^\infty \delta\big(\omega-\nu_G(\omega,t)\big)
G_s(\widetilde{\omega},t)d\widetilde{\omega},\\
&T_s(\omega,t)=C_\psi^{-1}\int_0^\infty\delta\big(\omega-\nu_W(\omega,t)\big)
W_s(\widetilde{\omega},t)\frac{d\widetilde{\omega}}{\widetilde{\omega}},
\end{aligned}\\
C_g\equiv\frac{1}{2}\int_{-\infty}^{\infty}\hat{g}(\xi)d\xi=\pi g(0),\quad
C_\psi\equiv\frac{1}{2}\int_0^\infty \hat{\psi}^*(\xi)d\xi/\xi,
\end{gathered}
\end{equation}
where $\nu_G\equiv{\rm Im}\big[\frac{\partial_tG_s(\omega,t)}{G_s(\omega,t)}\big]$ and $\nu_W\equiv{\rm Im}\big[\frac{\partial_tW_s(\omega,t)}{W_s(\omega,t)}\big]$ are the instantaneous phase velocities of the WFT and WT, respectively. In practice, the frequency scale is discretized, so one calculates the SWFT and SWT as $V_s(\omega,t)$ and $T_s(\omega,t)$ already integrated over the corresponding frequency bin (see e.g.\ the discussion in \cite{Iatsenko:tfr}).

Figure \ref{fig:sswftex} shows SWFTs constructed from the WFTs depicted in Figs.\ \ref{fig:wftex} and \ref{fig:wftexnoise} (see also \cite{Chen:14} for a systematic analysis of the effects of different kinds of noise on performance of the SWT). Clearly, synchrosqueezed TFRs are very concentrated and visually appealing. However, it has been found \cite{Iatsenko:tfr} that they do not possess better time or frequency resolution, i.e.\ do not allow for better reconstruction of components that are close in frequency or have high time variability (as compared to the original WFT/WT). Thus, the synchrosqueezing just sums all the interferences and other complications present in the WFT/WT into a more compact frequency regions so that, even if the components appear more separated as a result, this does not mean that their parameters can be better estimated. In this respect the SWFTs/SWTs are somehow similar to the WFT/WT skeletons (the corresponding transforms with only their amplitude peaks left) which, although being perfectly concentrated, do not obviously possess better resolution properties than the respective WFTs/WTs; see \cite{Iatsenko:tfr} for a more detailed discussion of this issue.

Nevertheless, it still remains to be established whether or not synchrosqueezing provides any advantages in terms of curve extraction, i.e.\ whether the ``correct'' amplitude peak sequences can be identified more easily in the SWFT/SWT than in the original WFT/WT. In other words, the following question is to be addressed: will performing synchrosqueezing first and then applying curve extraction methods to the resultant SWFT/SWT give more accurate results than applying these methods directly to the original WFT/WT?

Evidently, the schemes developed for the WFT/WT can straightforwardly be applied for tracing ridge curves in the SWFT/SWT. Nothing qualitatively changes, except that now one uses the amplitude peaks of the synchrosqueezed transforms. However, an immediate and serious drawback of this approach is that, in contrast to the case of the WFT/WT, the peak amplitudes in the synchrosqueezed transforms are not universally proportional to the amplitudes of the corresponding components; instead, they are largely determined by the parameters of frequency discretization being used, as illustrated in Fig.\ \ref{fig:wftsswft}. Thus, it can be seen that, even if one component has a smaller amplitude than the other, it may still have a much higher peak in the SWFT (see Fig.\ \ref{fig:wftsswft}(a,c)).

\begin{figure*}[t]
\begin{center}
\includegraphics[width=0.32\linewidth]{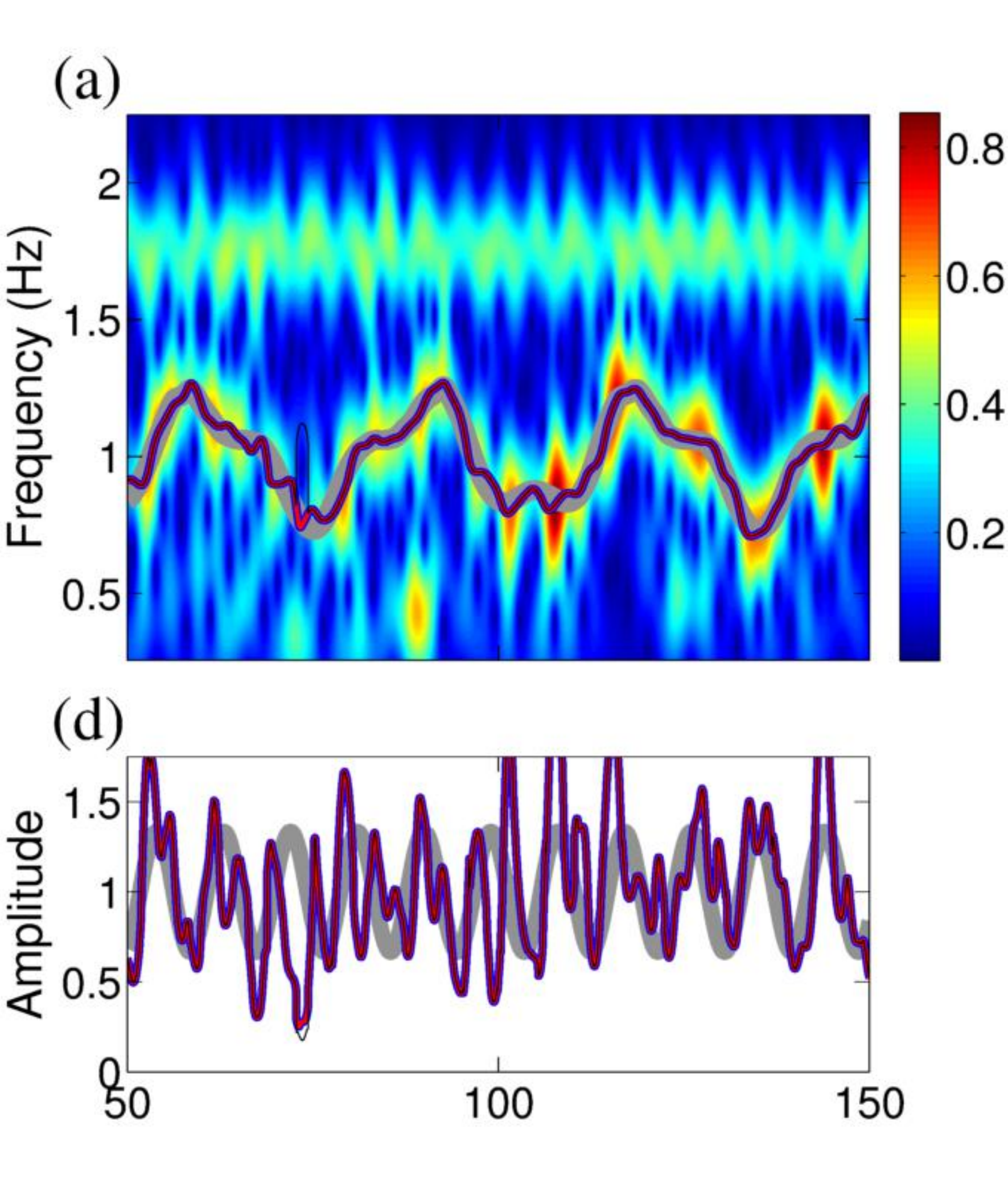}
\includegraphics[width=0.32\linewidth]{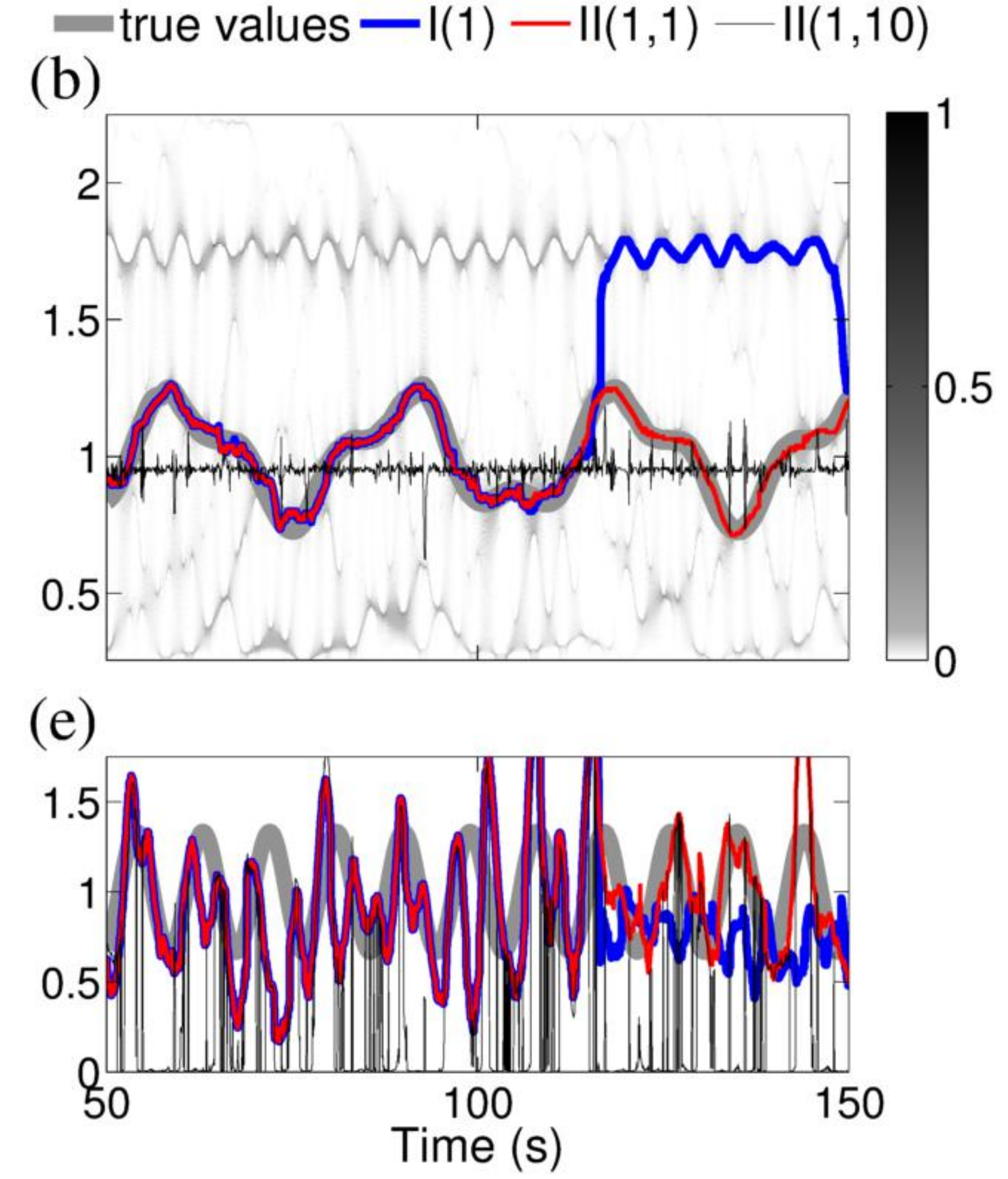}
\includegraphics[width=0.32\linewidth]{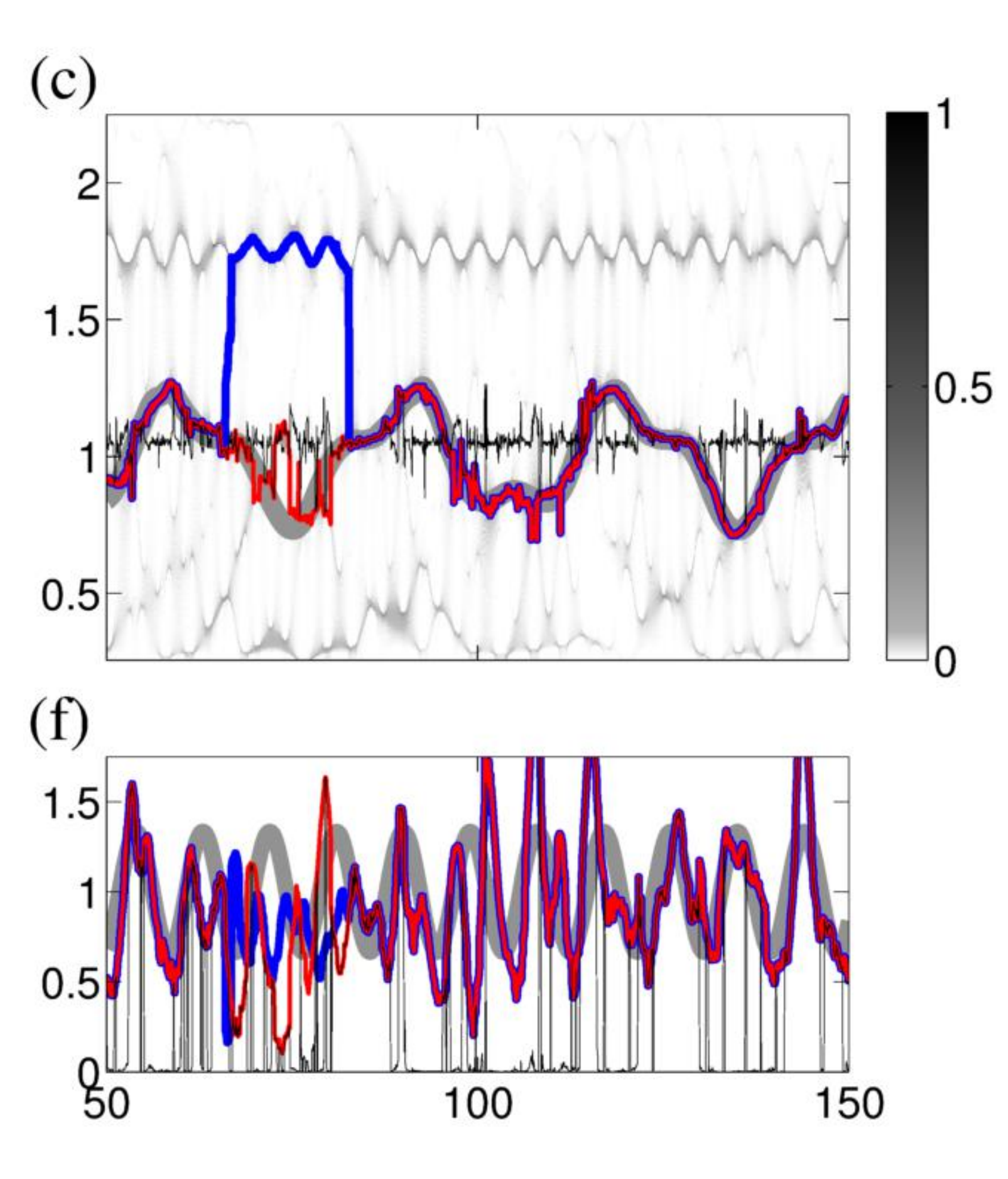}
\end{center}
\caption{Comparison of the curves extracted by different methods from the WFT (a) and SWFT (using the peaks (b) or integrated ridges (c)) for the first test signal (\ref{s1}) at a noise level $\sigma=0.6$ (\ref{signoise}). The lower panels (d,e,f) show the component's amplitude as reconstructed from the corresponding ridge curve by (\ref{ssppos}) for the SWFT, and in a similar manner (by integrating over the widest frequency regions of unimodal TFR amplitude around $\omega_p(t)$ at each time, see \cite{Iatsenko:tfr}) for the WFT.}
\label{fig:sscompare}
\end{figure*}

Generally, the relationships between the peaks will depend on the discretization step $\Delta\omega$, and this dependence proves to be highly nonlinear and time-varying, being influenced by many factors such as the instantaneous amplitude and frequency modulation of the component, its interference with other components and noise. Hence, the outcomes of different curve extraction methods when applied to synchrosqueezed transforms will also depend on the widths of frequency bins used. This effect is additionally augmented by the fact that, due to the non-smoothness of the SWFT/SWT, one cannot apply peak interpolation to better locate the ridges $\nu_m(t)$, so that they remain discrete, and such a discretization in turn affects performance of the extraction schemes (see Remark \ref{rem:deff}). Because of all these issues, the use of the SWFT/SWT peak amplitudes for discriminating between the components is in general not appropriate and can lead to unpredictable results, introducing considerable instability.

A possible way to avoid the drawbacks discussed above is to use the ``integrated'' ridges instead of the peaks. Thus, it is well-known \cite{Iatsenko:tfr,Daubechies:11,Auger:13} that in the case of the synchrosqueezed transforms the amplitude of the component should be estimated based on the overall sum of the SWFT/SWT over the (time-dependent) frequency region where it is concentrated. The problems attributable to use of the peaks can therefore be solved by using a more appropriate amplitude/frequency estimates. Hence, at each time $t$ we break the SWFT/SWT into the widest regions of non-zero amplitude $[\omega_-^{(m)}(t),\omega_+^{(m)}(t)]$. Then, instead of using peak values (\ref{ppos}), the ridge amplitudes $Q_m(t)$ and frequencies $\nu_m(t)$ (which are used in all procedures) are estimated from the corresponding regions as
\begin{equation}\label{ssppos}
\begin{gathered}
Q_m(t)=|x_m^{(a)}(t)|,\quad x_m^{(a)}(t)\equiv\int_{\omega_-^{(m)}(t)}^{\omega_+^{(m)}(t)} V_s(\omega,t)d\omega,\\
\nu_m(t)\equiv\operatorname{Re}\left[\big(x_m^{(a)}(t)\big)^{-1}\int \omega S_s(\omega,t)d\omega\right],
\end{gathered}
\end{equation}
for the SWFT, and similarly ($V_s(\omega,t)\rightarrow T_s(\omega,t)$) for the SWT. Since such $Q_m(t)$ do not depend on the widths of the frequency bins, being directly proportional to the true amplitudes of the corresponding components, while $\nu_m(t)$ now take continuous values, curve extraction methods based on integrated ridges are expected to give consistent results that are relatively unaffected by frequency discretization.

However, in both cases of using usual and integrated ridges, we have found most of the methods considered to perform either similarly, or often worse, if applied to the SWFT/SWT instead of the original WFT/WT. The corresponding results for the two test signals are shown in Supplementary Figs.\ 2, 3, 4 and 5; in all cases, the best performance was demonstrated by scheme II(1,1). Note that, in the case of weak frequency modulation (such as for the ECG signal), for some parameters the performance of the schemes might actually be slightly improved if using the SWFT/SWT peaks instead of the WFT/WT ones but, on the other hand, this causes the same schemes to fail completely for other parameters (see e.g.\ Supplementary Fig.\ 3). In any case, as discussed previously, the use of the SWFT/SWT peaks in the context of curve extraction is not generally appropriate.

A typical examples of the extracted curves are presented in Fig.\ \ref{fig:sscompare}, where one can see that, in contrast to the case of the WFT, the results of curve extraction from the SWFT become very sensitive to the method and its parameters being used. This is mainly because, in contrast to the usual WFT and WT, the synchrosqueezed transforms often contain a lot of ``spikes'' with small $Q_m(t)$ not corresponding to any component (see Fig.\ \ref{fig:wftsswft}(a,b)). These small peaks occur both due to noise and as a side effect of amplitude/frequency modulation or interference. Consequently, at any given time, there are numerous closely spaced candidate ridge points $\nu_m(t)$ in the SWFT/SWT, which makes it easier to switch between the curves corresponding to different components by building ``bridges'' between them (cf.\ blue lines in Fig.\ \ref{fig:sscompare}(a) and (b,c)), while for the WFT/WT this would require jumping large frequency distance in a sudden. Furthermore, this structure of the synchrosqueezed transforms allows selection of an almost straight curve formed mainly from the spurious ridges of close frequencies, and such a curve will indeed be returned if penalization of frequency or its time-derivative is strong enough (cf.\ black lines in Fig.\ \ref{fig:sscompare}(a) and (b,c)). Note that a similar situation would occur for the WFT/WT if we used all available frequencies as candidate ridge points $\nu_m(t)$, but using only the peaks (\ref{ppos}) avoids the corresponding drawbacks.

Finally, it should also be noted that the computational cost of curve extraction from the synchrosqueezed transforms is usually considerably higher than for the conventional smooth TFRs. Thus, the number of computations is proportional to the sum of squares of numbers $N_p(t)$ of ridge points $\nu_m(t)$ at each time (see Appendix), and these numbers are much larger for the SWFT/SWT than for the original WFT/WT.

\section{Limitations}\label{sec:limitations}

The methods proposed are subject to a few important limitations. First, all schemes are designed to extract accurately the curves corresponding to components that persist throughout the whole signal (or disappear only briefly). This is typically the case for signals of biological origin, such as recordings of ECG, EEG, respiration, or blood flow. On the other hand, when the signal contains transient components that are present only during short time frames, as is often the case e.g.\ in sound analysis, the curves returned by schemes I and II will most likely consist of the curves corresponding to different components appearing at similar frequencies but different times. This is because the proposed techniques do not have any built-in criteria to terminate curve extraction after a component ceases to exist. How best to formulate such a criterion is a separate topic, and will be the subject of future research.

Secondly, in common with virtually any curve extraction method, the proposed schemes can have problems with the signals containing components whose frequencies cross each other. In such cases it becomes unclear which path to follow after the crossing occurs. In practice one would like to select the profile which seems ``most consistent'', which in mathematical terms can be formulated as the most smooth. If the differences between components' amplitudes and/or frequency derivatives are high at the crossing point, then it is likely that the proposed schemes will return appropriate curves; otherwise they can generally select any path. Suppressing deviations of the higher derivatives of the component's frequency (in addition to the first one in both schemes) is likely to improve the situation, albeit with increased method complexity and computational cost.

Finally, it should be noted that to obtain reliable results with generally any method applied to the signal's TFR, the latter should represent appropriately at least the basic signal structure. How to achieve this is a general topic of time-frequency analysis (see e.g.\ \cite{Iatsenko:tfr,Mallat:08,Boashash:03}.

\section{Conclusions}\label{sec:conclusions}

We have developed and compared the techniques that can be used for ridge curve extraction from the WFT/WT, and discussed a number of related issues. Among the proposed approaches, scheme II($\alpha$,$\beta$) with $\alpha=\beta$ was shown to produce the best results. Its parameters $\beta$ and $\alpha$ control the strengths of suppression of the relative deviations of ridge frequency and its time-derivative from the corresponding median values, respectively. Although these parameters can be adjusted to better match any specific problem, due to high adaptivity of the approach the default choice $\alpha=\beta=1$ works well in the majority of cases (within the limitations discussed in the previous section). Thus, scheme II(1,1) appears to be of almost universal utility, being a type of ``just apply'' method that does not require any tuning by the user. The corresponding MatLab codes, as well as other useful time-frequency analysis tools, are freely available at \cite{freecodes}.

We have also tested the effects of synchrosqueezing \cite{Daubechies:11,Thakur:11,Thakur:13,Auger:13} in relation to curve extraction, and found that its drawbacks heavily outweigh the advantages. Thus, although scheme II(1,1) still remains the best and works reasonably well if applied to the synchrosqueezed transforms, in general the structure of the SWFT/SWT seems to be less suitable for curve extraction compared to that of the WFT/WT, at least for the methods considered.

\section*{Appendix: Fast path optimization of the functional with finite memory}\label{sec:conclusions}

Finding the solution $\omega_p(t)$ to the path optimization problem (\ref{pathopt}) is generally very expensive computationally, often being carried out by simulated annealing. However, if the functional $F[...]$ has finite memory, i.e.\ depends on the finite number of points selected at previous times (rather than the full history), then the optimal path can be found in O$(N)$ operations using dynamic programming techniques \cite{Bertsekas:95}. The corresponding algorithm is discussed in detail below.

Consider first the functional $F[Q_m(t_n),\nu_m(t_n),\omega_p(t_{n-1})]$, which depends only on the ridge point at the current time $t_n$ (characterized by $Q_m(t_n)$ and $\nu_m(t_n)$) and the frequency of the previous one $\omega_p(t_{n-1})$. This is basically the case utilized in all schemes presented in this work. The optimization problem (\ref{pathopt}) consists of finding the sequence of ridge point indices $m_c(t_n)$ maximizing the integral of this functional over time:
\begin{equation}\label{po1}
\underset{\{m_1,m_2,...,m_N\}}{\operatorname{argmax}}\sum_{n=1}^NF\big[Q_{m_n}(t_n),\nu_{m_n}(t_n),\nu_{m_{n-1}}(t_{n-1})\big].
\end{equation}
The ridge curve is then recovered as $\omega_p(t_n)=\nu_{m_c(t_n)}(t_n)$.

It is clear that at each time $t_n$ for each ridge $\nu_m(t_n)$ there exists a history of previous peaks $\{\widetilde{m}_c(m,t_n,t_1),\dots,\widetilde{m}_c(m,t_n,t_{n-1})\}$ which maximizes the integral to this point
\begin{equation}\label{po2}
\begin{aligned}
&U(m,t_n)=F[Q_m(t_n),\nu_m(t_n),\nu_{\widetilde{m}_c(m,t_n,t_{n-1})}(t_{n-1})]\\
&+\sum_{i=1}^{n-1}F[Q_{\widetilde{m}_c(m,t_n,t_i)},\nu_{\widetilde{m}_c(m,t_n,t_i)}(t_i),\nu_{\widetilde{m}_c(m,t_n,t_{i-1})}(t_{i-1})].
\end{aligned}
\end{equation}
What makes a fast path optimization possible is that, for functionals depending only on the current and previous points, if the profile $\{m_c(t)\}$ maximizing (\ref{pathopt}) includes $\nu_m(t_n)$, then it should include the best path to $\nu_m(t_n)$ as well: $\{m_c(t_1),\dots,m_c(t_{n})\}=\{\widetilde{m}_c(m,t_n,t_1),\dots,\widetilde{m}_c(m,t_n,t_{n-1}),m\}$. This is because the behavior of $m_c(t_{i=n+1,..,N})$ does not influence the integral over the previously extracted points $m_c(t_{i=1,..,n-1})$. Therefore, at each step we can leave only the best paths to each peak $\nu_m(t)$ and discard all the others.

It is useful to express $\widetilde{m}_c(m,t_n,t_i)$ through the matrix $q(m,t_n)$ which maps the peak number $m$ at time $t_n$ to the previous peak number in such a way that (\ref{po2}) is maximized. We therefore introduce
\begin{equation}\label{pnn}
\begin{gathered}
q[i](m,t_n)\equiv\widetilde{m}_c(m,t_n,t_{n-i})=q(q[i-1](m,t_n),t_{n-i+1}):\\
\begin{aligned}
q[0](m,t_n)&=m,\\
q[1](m,t_n)&=q(m,t_n)=\widetilde{m}_c(m,t_{n-1}),\\
q[2](m,t_n)&=q(q(m,t_n),t_{n-1})=\widetilde{m}_c(m,t_{n-2}),\\
&\dots
\end{aligned}
\end{gathered}
\end{equation}
What remains is to find at each time $t_n$ (starting from $t_1$), and for each ridge $m=1,\dots,N_p(t_n)$, the maximum value $U(m,t_n)$ of the integral up to this point and the index of the previous ridge $q(m,t_n)$ for which this maximum is achieved:
\begin{equation}\label{mpmn}
\begin{aligned}
&\mbox{\texttt{for $n=1,...,N$ and $m=1,...,N_p(t_n)$ do:}}\\
&
\begin{aligned}
q(m,t_n)=&
\begin{array}[t]{@{}r@{}l@{}}
\underset{k}{\operatorname{argmax}}\big\{ & F[Q_m(t_n),\nu_m(t_n),\nu_{k}(t_{n-1})]\\
                                          & +U(k,t_{n-1})\big\},\\
\end{array}\\
U(m,t_n)=&
\begin{array}[t]{@{}r@{}l@{}}
            & F[Q_m(t_n),\nu_m(t_n),\nu_{q(m,t_n)}(t_{n-1})]\\
\vphantom{\Big[} & +U(q(m,t_n),t_{n-1}),\\
\end{array}\\
\end{aligned}
\end{aligned}
\end{equation}
Then $U(m,t_N)$ represents the full integral (\ref{po1}) to each of the last ridges $\nu_m(t_N)$, and one has $m_c(t_N)=\operatorname{argmax}_mU(m,t_N)$, with the sequence corresponding to this index being the optimal path: $\{m_c(t)\}=\{q[N-1](m_c(t_N),t_N),\dots,q[1](m_c(t_N),t_N),m_c(t_N)\}$.

For example, for the functional $F[...]=\log Q_m(t_n)+w(\nu_m(t_n)-\omega_p(t_{n-1}),\alpha)$ (scheme I (\ref{es1})), we calculate
\begin{equation}\label{mpmniter}
\begin{aligned}
&t_1:\;\mbox{\texttt{for $m=1,...,N_p(t_1)$}}\\
&q(m,t_1)=0,\;U(m,t_1)=\log Q_m(t_1),\\
\midrule
&t_2:\;\mbox{\texttt{for $m=1,...,N_p(t_2)$}}\\
&q(m,t_2)=\underset{k}{\operatorname{argmax}}\big\{\log Q_m(t_2)+w(\nu_m(t_2)-\nu_k(t_1),\alpha)\\
&\hphantom{q(m,t_2)=\underset{k}{\operatorname{argmax}}\big\{}+U(k,t_1)\big\},\\
&U(m,t_2)=\log Q_m(t_2)+w(\nu_m(t_2)-\nu_{q(m,t_1)}(t_1),\alpha)\\
&\hphantom{U(m,t_2)=}+U(q(m,t_2),t_1),\\
\midrule
&t_3:\;\mbox{\texttt{for $m=1,...,N_p(t_3)$}}\\
&q(m,t_3)=\underset{k}{\operatorname{argmax}}\big\{\log Q_m(t_3)+w(\nu_m(t_3)-\nu_k(t_2),\alpha)\\
&\hphantom{q(m,t_2)=\underset{k}{\operatorname{argmax}}\big\{}+U(k,t_2)\big\},\\
&U(m,t_3)=\log Q_m(t_3)+w(\nu_m(t_3)-\nu_{q(m,t_2)}(t_2),\alpha)\\
&\hphantom{U(m,t_3)=}+U(q(m,t_3),t_2),\\
\midrule
&...
\end{aligned}
\end{equation}
where $q(m,t_1)$ is set to zero because there are no peaks before the starting time $t_1$.

Numerically, the $q(m,t_n)$ and $U(m,t_n)$ represent $M_p\times N$ matrices, updated at each step, where $M_p=\max_n N_p(t_n)$ is the maximum number of peaks; the excess entries $q(\{N_p(t_n)+1,..,M_p\},t_n)$ and $U(\{N_p(t_n)+1,..,M_p\},t_n)$ are set to Not-a-Numbers (NaNs). Since at each time $t_n$ we need to calculate for each of the $N_p(t_n)$ peaks the functional with each of the $N_p(t_{n-1})$ of the previous peaks (to find the one maximizing it), the overall computational cost of the procedure is O$(M_p^2 N)$ (or, more precisely, O$(\langle N_p(t_i)N_p(t_{i+1})\rangle N)$). The outcome of the algorithm is illustrated below on a schematic example:\newline\newline
\includegraphics[width=1.0\linewidth]{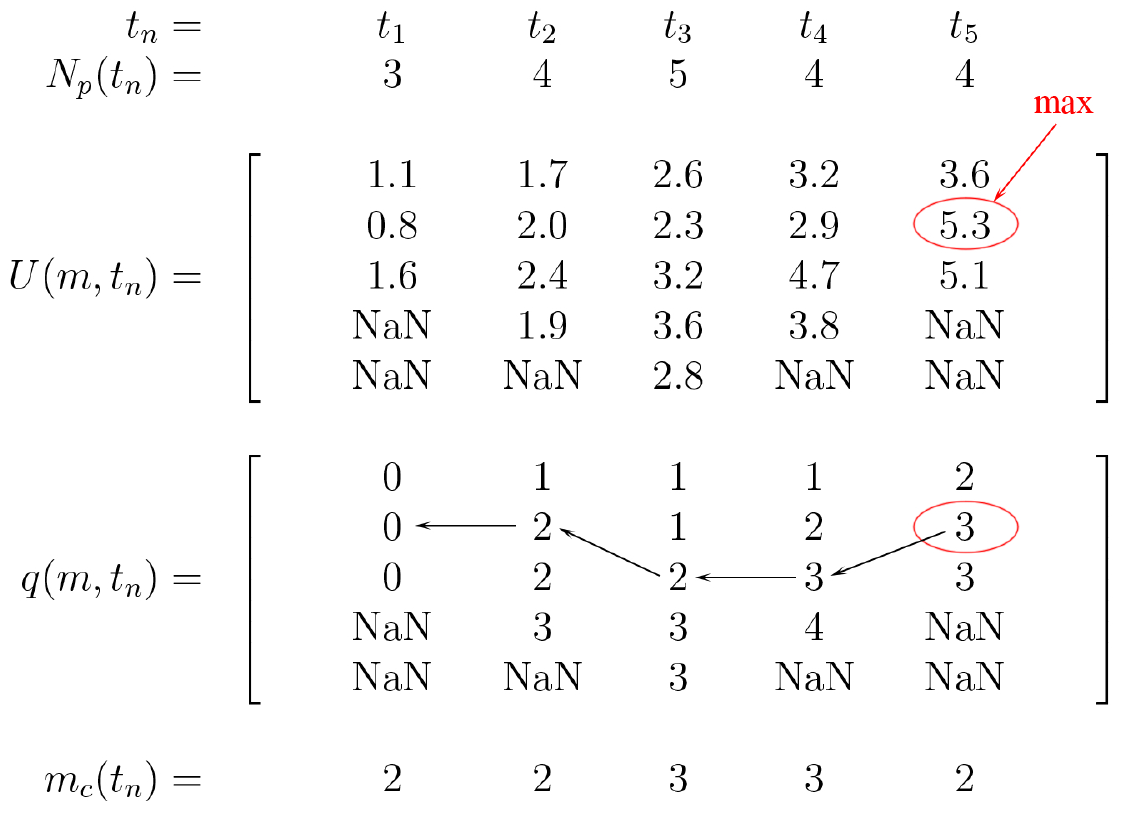}\\
Note, that in this example there are two ways of going from the second peak at time $t_1$: either to the second row ($m_c(t_2)=2$), corresponding to $U(2,t_2)=2.0$, or to the third one, corresponding to $U(3,t_2)=2.4$. The one-step scheme (\ref{onestepopt}) would select the third peak, but using the path optimization scheme we explore all the possibilities, and find out that going through the second one leads at the end to the higher path functional (\ref{pathopt}).

The path optimization for functionals depending on any finite number of previous peak positions (and not only one, as in (\ref{po1})) can be performed in a manner analogous to that outlined above. For example, if functional $F[...]$ depends on two previous points $\omega_p(t_{n-1})$ and $\omega_p(t_{n-2})$, then one will need to apply the same procedure but instead of single ridges treat their one-step sequences. Thus, in this case at time $t_n$ one selects the trajectory maximizing the path functional (\ref{pathopt}) to each of the $N_p(t_{n-1})\times N_p(t_n)$ point combinations $\{\nu_{k}(t_{n-1}),\nu_{m}(t_n)\}$. The general case of accounting for $d$ previous points is qualitatively similar, so the computational cost of the procedure is O$(M_p^{d+1}N)$.

\bibliographystyle{ieeetran}
\bibliography{ecurvebib}

\newpage\setcounter{figure}{0}
\begin{table*}[t]
  \centering
  \begin{tabular}{c}
  \textbf{\huge Supplementary Information}
  \end{tabular}
\end{table*}

\begin{figure*}[t!]
\includegraphics[width=1.0\linewidth]{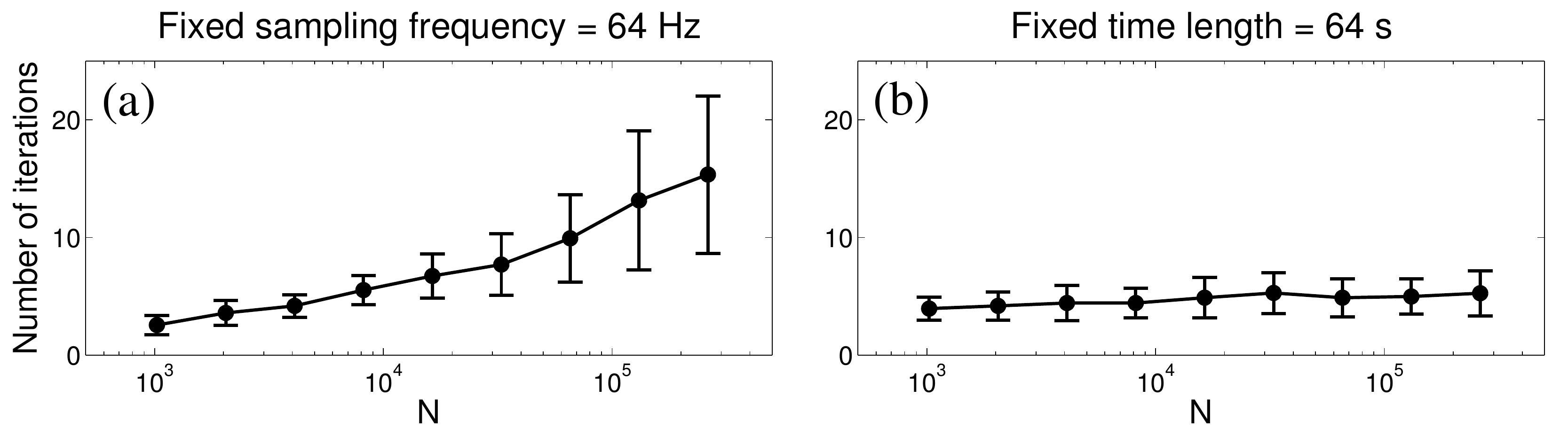}\\
\caption{The dependence on $N$ (the signal's length in samples) of the mean number of iterations needed for scheme II(1,1) to converge exactly in the case of a white noise signal; the error bars show $\pm 1$ standard deviation, and 40 independent noise realizations were used for each $N$. (a): The value of $N$ is varied by changing the time length of the signal. (b): The value of $N$ is varied by changing the sampling frequency of the signal. In all cases, scheme II was applied to the signals' WFTs calculated for the frequency range $[0,5]$ Hz, but the results do not change qualitatively if one calculates these WFTs for all available ranges (up to the Nyquist frequency).}
\end{figure*}

\begin{figure*}[t!]
\includegraphics[width=1.0\linewidth]{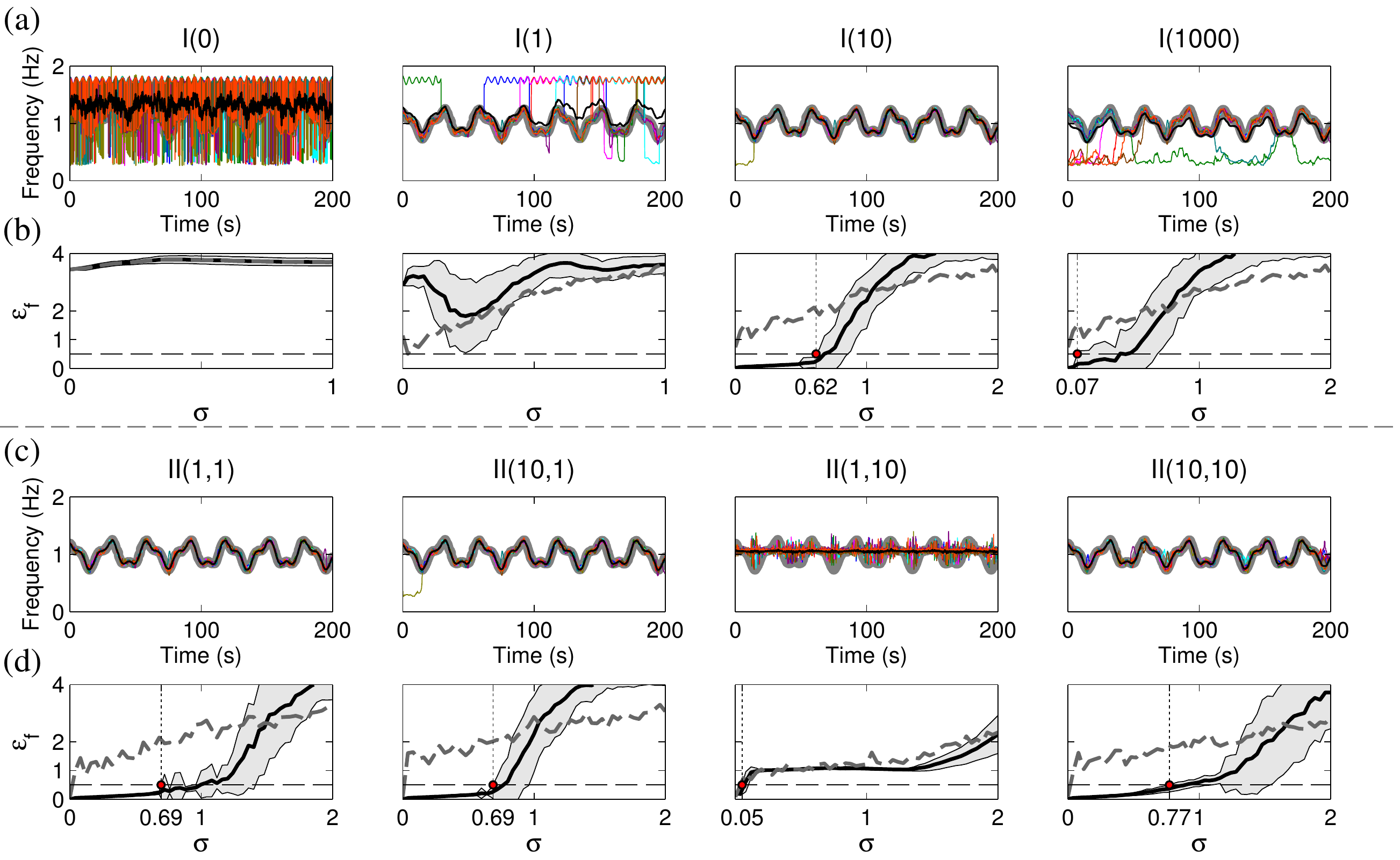}\\
\caption{Same as Fig.\ 3 in the manuscript, but the curve is extracted from the synchrosqueezed WFT using amplitude peaks as ridge points.}
\end{figure*}

\begin{figure*}[t!]
\includegraphics[width=1.0\linewidth]{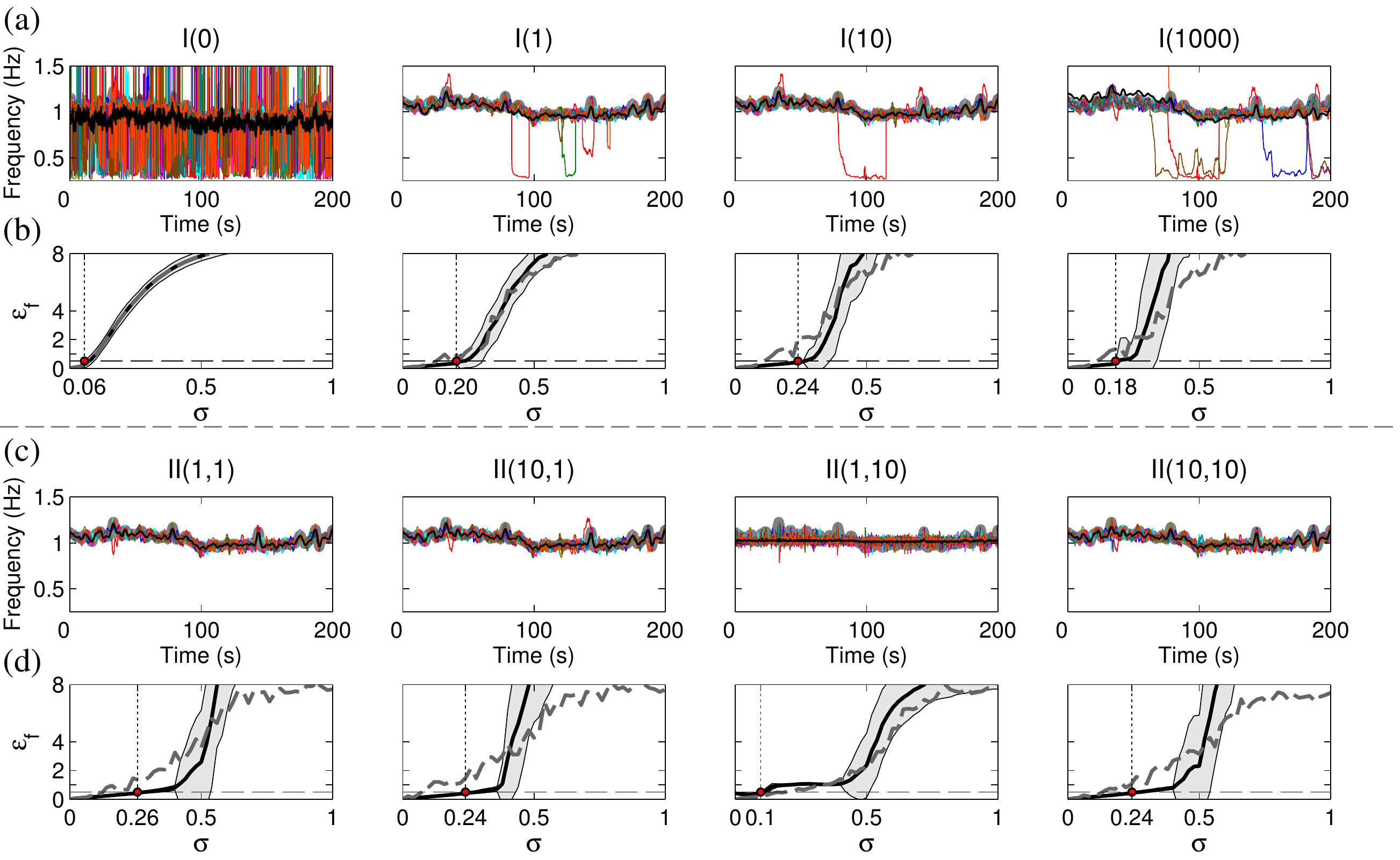}\\
\caption{Same as Fig.\ 4 in the manuscript, but the curve is extracted from the synchrosqueezed WFT using amplitude peaks as ridge points.}
\end{figure*}

\begin{figure*}[t!]
\includegraphics[width=1.0\linewidth]{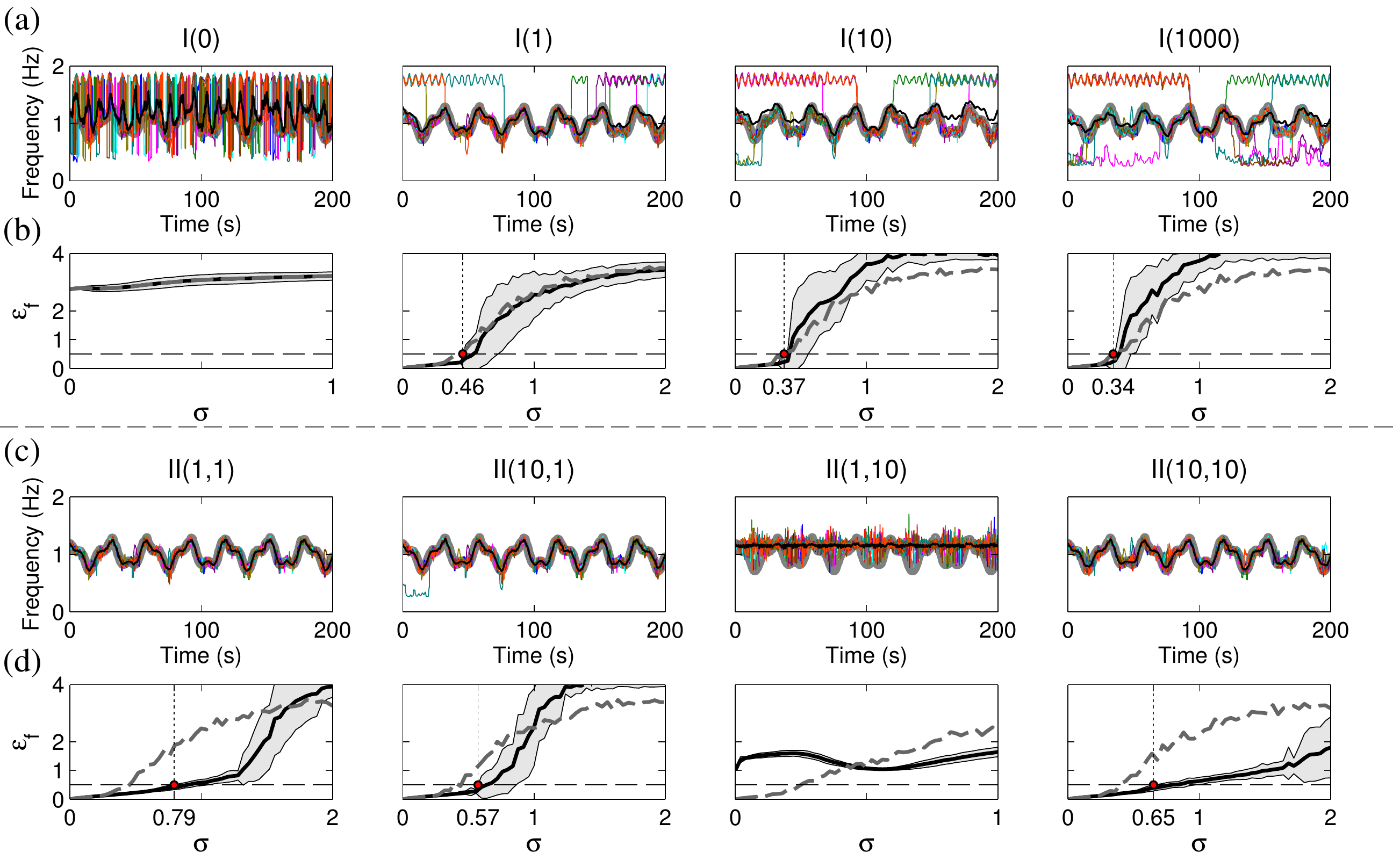}\\
\caption{Same as Fig.\ 3 in the manuscript, but the curve is extracted from the synchrosqueezed WFT using ``integrated'' ridge points (see Eq.\ (V.2) in the manuscript).}
\end{figure*}

\begin{figure*}[t!]
\includegraphics[width=1.0\linewidth]{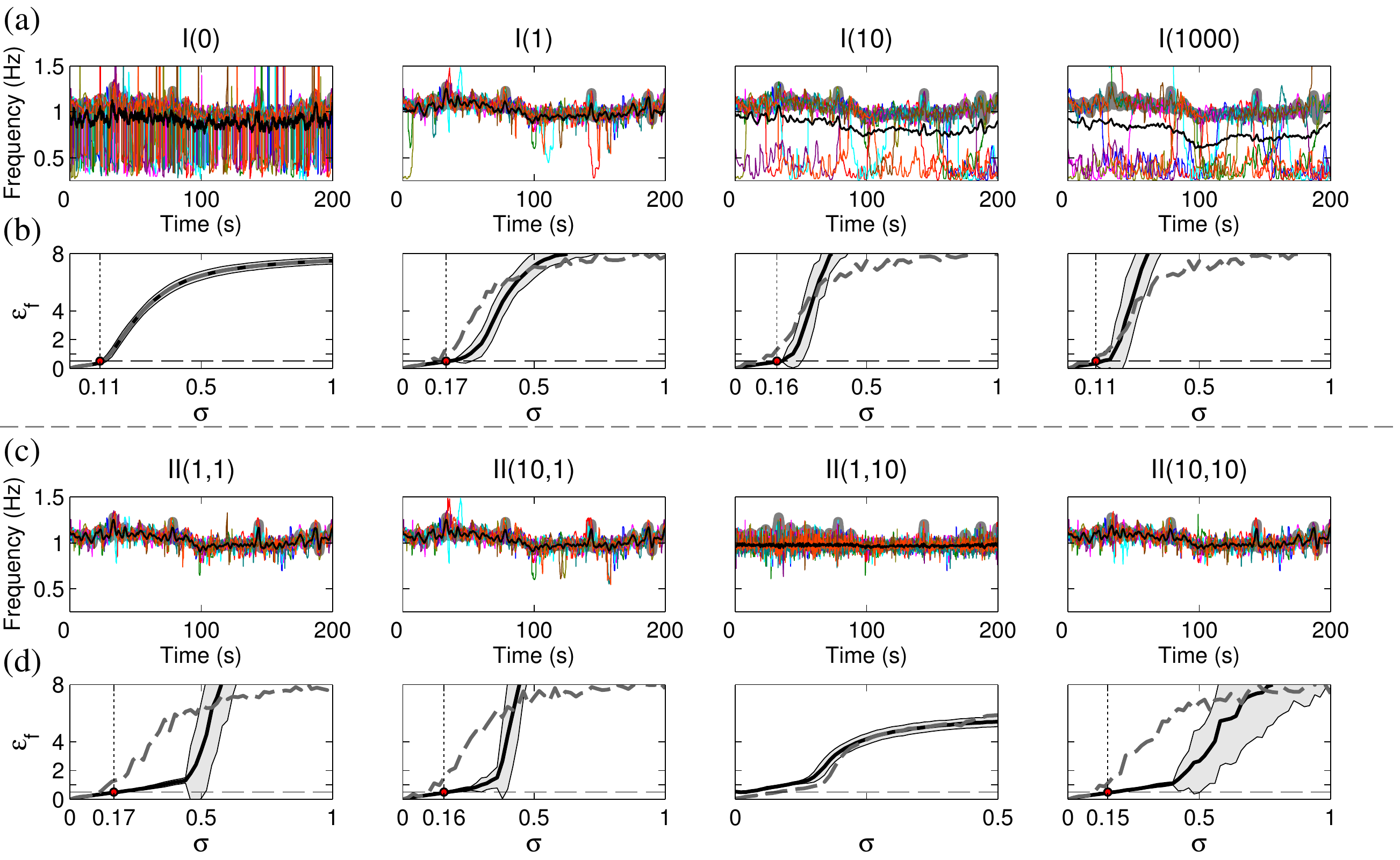}\\
\caption{Same as Fig.\ 4 in the manuscript, but the curve is extracted from the synchrosqueezed WFT using ``integrated'' ridge points (see Eq.\ (V.2) in the manuscript).}
\end{figure*}

\end{document}